\documentclass[12pt]{article} 
\usepackage{epsfig}
\usepackage{amsmath}
\usepackage{amsfonts}
\usepackage{latexsym}
\usepackage{amsmath,amssymb}
\usepackage{verbatim}
\usepackage{mathtools}
\usepackage{setspace}
\usepackage[all]{xypic}

\usepackage{tikz}
\usepackage[dvipdfm,hypertex]{hyperref}
\usepackage[textheight=9in, textwidth=6.5in, letterpaper]{geometry}

\numberwithin{equation}{section}

\def\d{{\frac{d}{d \rho}}}
\def\La{{L_{11}(\rho)}}
\def\Lb{{L_{12}(\rho)}}
\def\LLa{{L_{21}(\rho)}}
\def\LLb{{L_{22}(\rho)}}
\def\LLc{{L_{23}(\rho)}}
\def\LLd{{L_{24}(\rho)}}
\def\Laa{{L_{31}(\rho)}}
\def\Lbb{{L_{32}(\rho)}}
\def\Lcc{{L_{33}(\rho)}}
\def\Ldd{{L_{34}(\rho)}}
\def\Lc{{L_{13}(\rho)}}
\def\Rb{{R_{11}^{(2)}(\rho)}}
\def\Rc{{R_{12}^{(2)}(\rho)}}
\def\Rd{{R_{13}^{(2)}(\rho)}}
\def\Re{{R_{14}^{(2)}(\rho)}}
\def\LLe{{L_{25}(\rho)}}
\def\RRb{{R_{21}^{(2)}(\rho)}}
\def\RRc{{R_{22}^{(2)}(\rho)}}
\def\RRd{{R_{23}^{(2)}(\rho)}}
\def\RRe{{R_{24}^{(2)}(\rho)}}
\def\Lee{{L_{35}(\rho)}}
\def\Rbb{{R_{31}^{(2)}(\rho)}}
\def\Rcc{{R_{32}^{(2)}(\rho)}}
\def\Rdd{{R_{33}^{(2)}(\rho)}}

\newcommand{\bea}{\begin{eqnarray}}
\newcommand{\eea}{\end{eqnarray}}
\newcommand{\be}{\begin{equation}}
\newcommand{\ee}{\end{equation}}
\newcommand{\ba}{\begin{align}}
\newcommand{\ea}{\end{align}}

  \makeatletter
  \let\over=\@@over \let\overwithdelims=\@@overwithdelims
  \let\atop=\@@atop \let\atopwithdelims=\@@atopwithdelims
  \let\above=\@@above \let\abovewithdelims=\@@abovewithdelims

\newdimen\tableauside\tableauside=1.0ex
\newdimen\tableaurule\tableaurule=0.4pt
\newdimen\tableaustep
\def\phantomhrule#1{\hbox{\vbox to0pt{\hrule height\tableaurule width#1\vss}}}
\def\phantomvrule#1{\vbox{\hbox to0pt{\vrule width\tableaurule height#1\hss}}}
\def\sqr{\vbox{%
  \phantomhrule\tableaustep
  \hbox{\phantomvrule\tableaustep\kern\tableaustep\phantomvrule\tableaustep}%
  \hbox{\vbox{\phantomhrule\tableauside}\kern-\tableaurule}}}
\def\squares#1{\hbox{\count0=#1\noindent\loop\sqr
  \advance\count0 by-1 \ifnum\count0>0\repeat}}
\def\tableau#1{\vcenter{\offinterlineskip
  \tableaustep=\tableauside\advance\tableaustep by-\tableaurule
  \kern\normallineskip\hbox
    {\kern\normallineskip\vbox
      {\gettableau#1 0 }%
     \kern\normallineskip\kern\tableaurule}%
  \kern\normallineskip\kern\tableaurule}}
\def\gettableau#1 {\ifnum#1=0\let\next=\null\else
  \squares{#1}\let\next=\gettableau\fi\next}

\def\Or[#1]{{\text{O}}\left({#1}\right)}
\def\dotl[#1,#2]{\left\langle #1, #2 \right\rangle}
\def\dotlb[#1,#2]{[ #1, #2 ]}
\def\dotp[#1,#2]{(#1) \cdot (#2)}
\def\aff[#1,#2]{\hat{#1}(#2)}
\def\n4sym{{\cal N}=4 SYM}
\def\>{\rangle}
\def\<{\langle}
\def\weight[#1,#2,#3]{\{(#1),#2,#3\}}
\def\ads[#1]{$\text{AdS}_{#1}$}

\begin{document}

\begin{center}

{ \LARGE {\bf  Homogeneous Relaxation at Strong Coupling\\ 
\vspace{.3cm}
from Gravity}}

\vspace{0.5cm}

Ramakrishnan Iyer$^a$, and Ayan Mukhopadhyay$^b$
\vspace{0.3cm}

{\it $^a$ Department of Physics and Astronomy, University
of Southern California\\
Los Angeles, California 90089-0484, USA \\ 
\tt{\small ramaiyer@usc.edu}
}

\vspace{0.3 cm}

{\it $^b$ Laboratoire de Physique Th\'eorique et Hautes
Energies (LPTHE)\\
Universit\'e Pierre et Marie Curie -- Paris 6; CNRS UMR
7589 \\ 
Tour 13-14, 4$^{\grave{e}me}$ \'etage, Boite 126, 4 Place
Jussieu, 75252 Paris Cedex 05, France \\
\tt{\small ayan@lpthe.jussieu.fr}
}

\vspace{0.5cm}

\end{center}

\begin{abstract}
Homogeneous relaxation is a ubiquitous phenomenon in
semiclassical kinetic theories where the quasiparticles are
distributed uniformly in space, and the equilibration involves
only their velocity distribution. For such solutions, the
hydrodynamic variables remain constant. We construct
asymptotically AdS solutions of Einstein's gravity dual to such
processes at strong coupling, perturbatively in the amplitude
expansion, where the expansion parameter is the ratio of the
amplitude of the non-hydrodynamic shear-stress tensor to the
pressure. At each order, we sum over all time derivatives
through exact recursion relations. We argue that the metric has
a regular future horizon, order by order in the amplitude
expansion, provided the shear-stress tensor follows an equation
of motion. At the linear order, this equation of motion
implies that the metric perturbations are composed of zero wavelength quasinormal modes. Our method allows us to calculate the
non-linear corrections to this equation perturbatively in the
amplitude expansion. We thus derive a special case of our
previous conjecture on the regularity condition on the boundary
stress tensor that endows the bulk metric with a regular future
horizon, and also refine it further. We also propose a new outlook for heavy-ion phenomenology at RHIC and ALICE.

\end{abstract}

\pagestyle{empty}

\pagebreak
\setcounter{page}{1}
\pagestyle{plain}

\setcounter{tocdepth}{2}

\noindent
\tableofcontents

\section{Introduction}
The AdS/CFT correspondence \cite{Maldacena, Polyakov, Witten} has given us a framework to study non-equilibrium phenomena in gauge theories at strong 't Hooft coupling in
real time. For conformal gauge theories at strong 't Hooft coupling,
this correspondence, if applicable, implies that there exists a universal sector of
non-equilibrium states. This universal sector of states maps to five-dimensional spacetimes with metrics
which have regular future horizons, and are solutions of
Einstein's equation with a negative cosmological constant. 
Construction of such solutions of gravity,
perturbatively in the limit of slow spatial and temporal
variations, leads us to uncover purely hydrodynamic phenomena in
the dual gauge theory and also enables us to compute the
hydrodynamic transport coefficients systematically \cite{Policastro1, Policastro2, Janik1, Janik2,
Janik3, Baier, Bhattacharyya1, Natsuume}.

It is, of course, of theoretical and possibly experimental
interest, to uncover a wider class of non-equilibrium phenomena
beyond hydrodynamics through the gauge/gravity correspondence. The universal sector
of non-equilibrium states itself includes a huge spectrum of such 
non-equilibrium phenomena, including probably early time
evolution of the quark-gluon plasma, as has recently been studied for boost
invariant flows \cite{Peschanski}. \footnote{For related work with sources for field theory operators like metric perturbations turned on, please also see \cite{Chesler1}.}

The universal sector has a special characteristic in that all
states constituting it can be uniquely characterized and their
dynamics can be completely determined by the expectation
value of the energy-momentum tensor alone \cite{myself1}. This follows from the
dual gravity description. In an earlier work \cite{myself2}, we have proposed a field-theoretic explanation by
drawing analogy to \textit {conservative solutions} of the
Boltzmann equation that we have constructed, and briefly explain below. 

It can be shown that an appropriate
relativistic semiclassical Boltzmann equation captures all
perturbative non-equilibrium phenomena in non-Abelian gauge
theories at sufficiently high temperature \cite{Arnold1, Arnold2}. The conservative
solutions of the relativistic semicalssical Boltzmann equation exist for all values of the 't Hooft coupling and rank of the gauge group. These \textit{special} solutions can again be completely
characterized and their dynamics can be completely
determined by the energy-momentum tensor alone. 

In these solutions, the components of the energy-momentum tensor follow a closed set of equations of motion which can be derived systematically from the Boltzmann
equation. In addition to the equations for
conservation of energy and momentum, this set also
contains equations for the evolution of the shear-stress tensor.
The other parameters of the quasiparticle distribution function,
like the heat current for instance, do not decouple, but are
\textit{algebraically} determined by the energy-momentum tensor
and their spatial derivatives in a local inertial frame where the mean velocity of the quasiparticles vanishes, and thus have no independent 
dynamical parts. 

The dynamics is determined by the energy-momentum tensor alone,
because any solution to the closed set of equations of motion of
the energy-momentum tensor can be lifted to a unique solution of
the Boltzmann equation through the other algebraically determined parameters. 

A special class of solutions to these
equations are purely hydrodynamic in nature and are known as
\textit{normal solutions} in the literature. \footnote{The \textit{normal} solutions
were first found by Enskog \cite{Enskog} in order to provide a systematic way of
calculating transport coefficients from the Boltzmann equation. These solutions 
can be determined exactly by the hydrodynamic variables, as all other parameters
of the quasiparticle distribution are determined algebraically by the hydrodynamic variables and their spatial derivatives, in a local inertial frame where the mean velocity of the quasiparticles vanishes. Many formal aspects of these solutions were 
clarified in \cite{Burnett, Chapman}, and they were found for the relativistic
semiclassical Boltzmann equation by Stewart \cite{Stewart}.} Further, any
solution of the Boltzmann equation can be approximated by an
appropriate \textit{conservative solution} at sufficiently late
times \cite{myself2}.

We proposed that the conservative solutions, when extrapolated to
strong 't Hooft coupling and large rank of the gauge group,
exhibit universality in their dynamics and constitute the
states of the universal sector. This explains why all states in the
universal sector can be determined by the energy-momentum tensor 
alone.

This proposal amounts to saying
that the dual solutions of pure gravity will have regular future
horizons, provided the boundary energy-momentum tensor follows a
closed set of equations of motion. The constraints of the equations
of motion of gravity imply conservation of energy and momentum. However these do not suffice to ensure regularity of the future horizon in the bulk, or determine the evolution of the boundary energy-momentum tensor completely.  Our proposed set of equations also includes
the equations of motion for the shear-stress tensor, which determine the evolution of the boundary energy-momentum tensor completely and ensure regularity of the future horizon in the bulk for the right values of the phenomenological parameters.

The structure of these equations cannot be related to the
corresponding structure in the case of conservative solutions
because we do not have a known
kinetic description at strong coupling. However, if we know the
hydrodynamic transport coefficients up to some orders, we can
phenomenologically construct the most general equation of motion
of the shear-stress tensor in appropriate expansion parameters
as we will discuss later. \footnote{Our study of homogeneous relaxation shows that the amplitude expansion of these equations of motion for the shear-stress tensor was not done correctly in \cite{myself2}. It was actually misguided by the structure of the amplitude expansion in the Boltzmann limit. Consequently,  we failed to connect our equation with the quasinormal modes after linearization. We will give the correct general phenomenological equation here.} This equation has purely
hydrodynamic solutions, just like \textit{conservative solutions}, 
but it also has more general solutions
corresponding to non-hydrodynamic relaxation.

The aim of this work is to chalk out a course for derivation of equations of motion for the energy-momentum tensor,
directly from the requirement of regularity of the
future horizon in the gravity dual, for a special case of non-hydrodynamic relaxation. This
special case is homogeneous relaxation. We
will concentrate on the homogeneous case for two important
reasons, firstly because they provide the simplest instances of
purely non-hydrodynamic approach to equilibrium in general, and
secondly because we can easily make a connection here with the
physics of non-hydrodynamic branches of quasinormal modes of
black branes.

Homogeneous relaxation to equilibrium is a feature of special
solutions of all kinetic theories. In the case of the Boltzmann
equation, for instance , these solutions correspond to the
quasiparticle distribution function being spatially uniform, but being
non-Maxwellian in velocity space. The higher moments of the
velocity distribution relax such that they vanish at
equilibrium, while the first five velocity moments 
corresponding to the hydrodynamic variables - 
the density, the mean velocity, and the temperature - remain constant.
\footnote{Spatially uniform solutions of the Boltzmann equation for special
intermolecular potentials were first found by T. Carleman \cite{Carleman}. These solutions
were also isotropic. More general solutions were later found by Wild \cite{Wild}
and Morgenstern \cite{Morgenstern}.}

In such solutions, the components
of the energy-momentum tensor in the globally defined inertial
frame with constant zero mean velocity, takes the form,
\begin{equation}\label{form1}
t_{00} = \epsilon , \quad t_{0i} = t_{i0} = 0, \quad t_{ij} =
p\delta_{ij} +\pi_{ij}(t),
\end{equation}
where $\epsilon$ is the energy density, $p$ is the pressure
dependent on the constant temperature $T$ and $\pi_{ij}(t)$ is
the shear-stress tensor. This form of the energy-momentum tensor
trivially satisfies conservation of energy and momentum,
$\partial^{\mu}t_{\mu\nu}=0$ (or equivalently the hydrodynamic
equations) for any $\pi_{ij}(t)$. We also require $\epsilon = 3p$ and
$\pi_{ij}\delta_{ij}=0$ so that the energy-momentum tensor is traceless.

A special class of such homogeneous solutions are also
conservative solutions, where the full solution can be
determined from the solution to the closed set of equations of
motion for $\pi_{ij}(t)$. These equations can be
systematically expanded in the amplitude expansion parameter, which is the
typical value of $\pi_{ij}$ divided by the constant pressure.

Remarkably, the scalar channel of quasinormal modes for AdS black branes, which have
zero wavelengths \cite{Horowitz, Starinets, Nunez, Kovtun, Berti} \footnote{For an earlier work please see \cite{KalyanaRama}.}, also suggest a similar form of the energy-momentum tensor. \footnote{The quasinormal modes of AdS black branes reproduces all the poles in the retarded Green's function of the dual guage theory \cite{Nunez, Kovtun}, so they carry information regarding linear thermal relaxation in the gauge theory.} We have a tower of such quasinormal modes, so that for such perturbations
the energy-momentum tensor takes the following form by gauge/gravity duality,
\begin{equation}\label{formb}
t_{\mu\nu}(t) = t_{\mu\nu}^{(0)} +
\displaystyle\sum\limits_{n=1}^{\infty}\left(a_{n \mu\nu}e^{-i\omega_{R_n}t}+a_{n
\mu\nu}^{*}e^{i\omega_{R_n}t}\right)e^{-\omega_{I_n}t},
\end{equation}
where $t_{\mu\nu}^{(0)}$ corresponds to the unperturbed AdS black brane and $\pm \ \omega_{R_n} - i\omega_{I_n}$ (with
$\omega_{I_n} > 0$) are the non-hydrodynamic overtones in the scalar channel. Also $a_{n\mu\nu}$ is traceless and purely spatial. Given that $t_{\mu\nu}^{(0)}$ is traceless,
it follows that $a_{n\mu\nu}$ is traceless as well. As the perturbations are in the scalar channel, $a_{n\mu\nu}$ is also transverse to the global boost four-vector $u^\mu$ of the black brane. It is not difficult to see that the form \eqref{formb} of the energy-momentum tensor is equivalent to \eqref{form1} in the global inertial frame where the boost four-vector is (1, 0, 0, 0).

Here, we will develop a systematic method to construct solutions
of gravity which have the boundary energy-momentum tensor of the
form (\ref{form1}), by summing over all time
derivatives, but perturbatively in the amplitude expansion parameter. \footnote{A class of  translationally invariant and isotropic solutions were previously constructed in \cite{Minwalla}, where the sources in the boundary corresponding to the non-normalizable modes were turned on and black hole formation in the bulk was studied. Our method of summing over time derivatives in the perturbation expansion here will be different and we can deal with arbitrary anisotropic configurations. We will adopt a different procedure for regularity analysis of the metric at the horizon here. For related earlier numerical study of homogeneous anisotropic asymptotically AdS metrics, please see \cite{Chesler2}.}

We construct the solution both in Fefferman-Graham coordinates
and the ingoing Eddington-Finkelstein coordinates. It will be
easier to construct the solution in Fefferman-Graham coordinates
first, because the constraints could be easily solved. The Fefferman-Graham
coordinates will also be suitable for achieving a general construction of metrics, uniform
in the spatial boundary coordinates.

We can then obtain the metric in ingoing Eddington-Finkelstein
coordinates by doing coordinate transformation systematically.
The coordinate transformation also receives corrections
order by order in the amplitude expansion, and at each order all time derivatives
can be summed up. The whole method will be, in fact, a
generalization of the method adopted in the purely hydrodynamic
case in \cite{myself1}.

The metric can be expected to be regular order by order in the amplitude expansion
for appropriate shear-stress tensor $\pi_{ij}(t)$. The shear-stress tensor becomes arbitrarily small close to equilibrium, analogous to the spacetime variations of the hydrodynamic variables. In the latter case, the metric is regular order by order in the derivative expansion for right choice of transport coefficients. Thus \emph{we can expect that the metric should be regular for appropriate $\pi_{ij}(t)$, order by order in the amplitude expansion after summing over all time derivatives at each order}. As we will see, the time derivatives of $\pi_{ij}(t)$ will not become arbitrarily small even close to equilibrium. This will necessitate summing over all time derivatives at each order in the amplitude expansion.

We can argue that the metric is manifestly regular in the ingoing Eddington-Finkelstein
coordinates order by order in the amplitude expansion when $\pi_{ij}(t)$ follows an equation
of motion. \emph{This equation of motion can also be found order by order in the amplitude expansion, summing over all time derivatives at each order}. At the first order, we reproduce the scalar channel of quasinormal modes at zero wavelengths. At the next order, we obtain non-linear corrections, which can potentially modify the behavior drastically at early times.

\textbf{Organization of the paper :} The organization of the rest of the paper is as follows. In
Section 2, we will develop the method to construct 
solutions of gravity uniform in the spatial boundary coordinates, in the Fefferman-Graham system,  order by order in the amplitude expansion while summing
over all time derivatives. In Section 3, we
will translate these solutions to the Eddington-Finkelstein
coordinates. In Section 4, we will analyze the regularity
condition on the future horizon and see how it gives us an
equation of motion for $\pi_{ij}(t)$. In Section 5, we will show how the analysis done here can be used to study the space-time evolution of the matter created by ultrarelativistic heavy-ion collisions at RHIC and ALICE. Finally in the Discussion we will conclude by mentioning a few open questions.

\section{The metric in Fefferman-Graham coordinate system in the
amplitude expansion}

In this section, we will develop a method to construct
asymptotically AdS metrics, where the energy-momentum tensor at
the boundary takes the same form \eqref{form1} as in homogeneous relaxation
in kinetic theories. The metric should have a regular future
horizon when this boundary energy-momentum tensor satisfies further
constraints, which will be investigated later.

We will construct the metric order by order in the amplitude
expansion, while summing over all time derivatives. We
will use Fefferman-Graham coordinate system for both, a purpose
of principle and a purpose of convenience.

The purpose of principle is that the Fefferman-Graham coordinate
system allows us to construct solutions for the most general
form of the energy-momentum tensor corresponding to homogeneous
relaxation, before we impose constraints for regularity at the
future horizon. A similar point of view was taken in
\cite{myself1} in the purely hydrodynamic case, where the metric
in the bulk was constructed in Fefferman-Graham coordinates for purely
hydrodynamic boundary energy-momentum tensor with arbitrary transport
coefficients, in the derivative expansion. It was then shown
that, for a unique choice of transport coefficients, the future
horizon was regular order by order in the derivative expansion.

In the present case of homogeneous relaxation, we have no
standard phenomenological equation for the shear-stress tensor that can be reliably applied at strong coupling. Thus, in principle,
we should construct the metric for the most general
\textit{form} of the shear-stress tensor at the boundary and then find out how
the regularity condition at the future horizon in the bulk constrains it.
When the boundary energy-momentum tensor is arbitrary, the
Fefferman-Graham coordinate system is the suitable choice for
constructing the metric.

The purpose of convenience is that the Fefferman-Graham
coordinate system allows us to satisfy
the constraints easily. The constraints can be elegantly solved
in the Fefferman-Graham coordinates, because they simply impose
boundary conditions on the dynamical equations for evolution of
the boundary metric in the radial direction. However, in other
coordinate systems, the dynamical equations are not simply the
evolution of the boundary metric. So, it is not easy to generate
a simple algorithm to satisfy the constraints. We will find that
the constraints here are more involved than in the purely
hydrodynamic case, so the Fefferman-Graham coordinate system is
indeed very convenient.

We will begin with obtaining the form of the metric order by
order in the amplitude expansion. Then we will show how we
obtain the equations for motion at each order in the amplitude
expansion and how we can sum over all time derivatives at each
order. The procedure will involve Fourier transformation of the
time dependence, and this will introduce a few subtleties
which we will also discuss.

\subsection{The form of the metric in the amplitude expansion}

Einstein's equation of gravity in five dimensions with a
negative cosmological constant takes the following convenient
form
\begin{equation}
R_{MN}- \frac{1}{2}RG_{MN}=\frac{6}{l^2}G_{MN}.
\end{equation}
Any metric which solves the above equation will have constant
scalar curvature $R = -4/l^2$. Further, any asymptotically AdS
metric which is a solution of this equation can be written in
the Fefferman-Graham coordinate system in the form,
\begin{equation}\label{FG}
ds^2=
\frac{l^2}{\rho^2}\left(d\rho^2+g_{\mu\nu}(\rho,z)dz^{\mu}dz^{\nu}\right).
\end{equation}
This coordinate system should be able to cover a part of the the
five-dimensional upper half plane $\rho\geq 0$, so it will have
a coordinate or a real singularity only at a finite radial
distance from the boundary $\rho = 0$ \cite{Fefferman, myself1}.
For the rest of this paper, we will choose our units such that
$l=1$.

The boundary metric is defined as
\begin{equation}
g_{(0)\mu\nu}(z)= \lim_{\rho\rightarrow 0}g_{\mu\nu}(\rho,z).
\end{equation}
By rules of gauge/gravity duality, which we will not review
here, it turns out that the boundary metric is the metric on the
four dimensional spacetime in which the gauge theory is living
\cite{Polyakov, Witten}.

When the boundary metric is flat, i.e. when $g_{(0)\mu\nu}(z) =
\eta_{\mu\nu}$, the four dimensional metric $g_{\mu\nu}(\rho,
z)$ in (\ref{FG}) will have a Taylor expansion in the radial
coordinate near $\rho = 0$ \cite{myself1}, such that
\begin{equation}\label{te}
g_{\mu\nu}(\rho, z) =\eta_{\mu\nu}
+\displaystyle\sum\limits_{n=0}^\infty
g_{(2n)\mu\nu}(z)\rho^{2n}.
\end{equation}
By the rules of gauge/gravity duality, when the boundary metric
is flat, it also turns out that $g_{(4)\mu\nu}$ is the
expectation value of the energy-momentum tensor in the dual
non-equilibrium state (aside from a factor which is essentially a
power of the rank of the gauge group \cite{Henningson,
Skenderis}). This result can also be obtained in a manifestly coordinate-independent
manner \cite{Balasubramanian, Skenderis}. Since we are in the limit where the
rank of the gauge group is infinite, we will normalize
$t_{\mu\nu}$, the expectation value of the energy-momentum
tensor in the dual state, such that $g_{(4)\mu\nu}(z) =
t_{\mu\nu}(z)$.

We note that this identification of $g_{(4)\mu\nu}$ wth
$t_{\mu\nu}$ makes sense only when the bulk metric has a regular
future horizon. More generally, we should think of it as a
boundary stress tensor.

Einstein's equation automatically guarantees
that $g_{(4)\mu\nu}$ satisfies the equation of conservation of
energy and momentum, $\partial^{\mu}g_{(4)\mu\nu}=0$, and is
also traceless, i.e. it satisfies $Tr (g_{(4)}) =0$
\cite{Henningson, Skenderis}. These allow us to identify it more
generally with a conformally covariant energy-momentum tensor
even when the bulk metric is not endowed with a regular future
horizon.

Einstein's equation for the metric (\ref{FG}) reduces to a
tensor, a vector and a scalar equation of motion. The tensor
equation, which gives the dynamical equation of motion for
evolution of the boundary metric in the radial coordinate, is
\begin{equation}\label{ed}
\frac{1}{2}g^{\prime\prime
}-\frac{3}{2\rho}g^{\prime}-\frac{1}{2}g^{\prime}g^{-1}g^{\prime}+\frac{1}{4}Tr(g^{-1}g^{\prime})g^{\prime}-\frac{1}{2\rho}Tr(g^{-1}g^{\prime})g-Ric(g)=0,
\end{equation}
where $'$ denotes differentiation with respect to the radial
coordinate $\rho$. The vector and tensor equations give the
constraints on $g_{(4)\mu\nu}$. The vector equation is
\begin{equation}\label{ecv}
\nabla_{\mu}Tr(g^{-1}g^{\prime})-\nabla^{\nu}g^{\prime}_{\mu\nu}
= 0 ,
\end{equation}
where $\nabla$ denotes the covariant derivative constructed from
$g$. The scalar equation is,
\begin{equation}\label{ecs}
Tr[g^{-1}g^{\prime\prime}]-\frac{1}{\rho}Tr[g^{-1}g^{\prime}]-\frac{1}{2}Tr[g^{-1}g^{\prime}g^{-1}g^{\prime}]
= 0.
\end{equation}

Actually, it turns out that the true dynamical equation is 
\begin{eqnarray}\label{tensor}
&&\frac{1}{2}g^{\prime\prime}-\frac{3}{2\rho}g^{\prime}-\frac{1}{2}g^{\prime}g^{-1}g^{\prime}+\frac{1}{4}Tr(g^{-1}g^{\prime})g^{\prime}-Ric(g)\nonumber\\&&+g\Big[\frac{1}{6}R(g)+\frac{1}{24}Tr(g^{-1}g^{\prime}g^{-1}g^{\prime})-\frac{1}{24}\left(Tr(g^{-1}g^{\prime})\right)^2\Big]=0,
\end{eqnarray}
which can be obtained by combining the tensor equation with the salar
equation, after multiplying the latter with the four-dimensional metric $g$ \cite{myself1}. Also, the true scalar constraint is
\begin{equation}\label{scalar}
R(g)+\frac{3}{\rho}Tr(g^{-1}g^{\prime})+\frac{1}{4}Tr(g^{-1}g^{\prime}g^{-1}g^{\prime})-\frac{1}{4}[Tr(g^{-1}g^{\prime})]^2=0,
\end{equation}
which can be obtained by combining the trace of the tensor equation with the scalar equation.

The best way to see how these equations work is to substitute
the form of the metric (\ref{te}) in (\ref{tensor}),
({\ref{ecv}) and (\ref{scalar}), and Taylor expand these
equations in $\rho$ about $\rho=0$ (the boundary). At
the leading order, the dynamical equation (\ref{tensor}) does
not determine $g_{(4)\mu\nu}$, but the vector constraint
(\ref{ecv}) imposes the conservation of energy and
momentum and the scalar constraint imposes the condition that
it should be traceless. This is expected because
$g_{(4)\mu\nu}$, being the boundary stress-tensor, is indeed 
independent data other than the boundary metric $g_{(0)\mu\nu}$
on the initial hypersurface $\rho=0$, so it cannot be determined
by the dynamical equation. On the other hand, the constraints
impose the desired conditions on this initial data.

Using the true dynamical equation (\ref{tensor}), we can
determine all the higher coefficients in the Taylor expansion
$g_{(2n)}$ for $n\geq 3$ uniquely and algebraically in terms of
polynomials of $g_{(4)\mu\nu}$ and its derivatives, or
equivalently in terms of $t_{\mu\nu}$ and its derivatives. A few
coefficients are
\begin{eqnarray}\label{coeffs}
g_{(6)\mu\nu}&=&-\frac{1}{12}\Box t_{\mu\nu}, \nonumber\\
g_{(8)\mu\nu}&=&\frac{1}{2}t_{\mu}^{\phantom{\mu}\rho}t_{\rho\nu}-\frac{1}{24}\eta_{\mu\nu}(t^{\alpha\beta}t_{\alpha\beta})+\frac{1}{384}\Box^2t_{\mu\nu},
\ ... \ ,
\end{eqnarray}
where all indices have been lowered or raised using the boundary
metric $\eta$ or its inverse. It is easy to see why in these
coefficients of the Taylor expansion (\ref{te}), the derivatives
come only in pairs. The only sources of these derivatives in the
dynamical eq. (\ref{tensor}) are $Ric(g)$ and $R(g)$ where they
do occur in pairs only.

When the coefficients of the Taylor expansion (\ref{te}) (like
the ones above in (\ref{coeffs})), as determined by the true
dynamical eq. (\ref{tensor}), are substituted in the vector
constraint (\ref{ecv}) and scalar constraint (\ref{scalar}), we
find that the constraints trivially vanish, provided
$\partial^{\mu}t_{\mu\nu} = 0$ and $Tr(t)=0$ (as obtained
at the lowest order). The consistency of the power series
expansion can be demonstrated once this trivialisation of the
constraint equations has been proved to all orders and this has
been done in \cite{myself1}.

The discussion above has been general and will apply to any
$t_{\mu\nu}$ as an initial data on the hypersurface $\rho=0$. We
will now specialize to the form of the energy-momentum tensor in
a non-equilibrium state undergoing homogeneous relaxation to
equilibrium.

To specify this form, we have to define the hydrodynamic
variables, the four velocity $u^{\mu}$ and the temperature $T$
first. In kinetic theories, usually $u^{\mu}$ is defined as the
local velocity of particle transport. However, it will
convenient for us to use the so-called Landau frame in which
$u^{\mu}$ is the local velocity of energy transport. By this
definition, $u^{\mu}$ is also a timelike unit vector satisfying
$u^{\mu}u_{\mu}= -1$. We will define the temperature such that
the local energy density given by $\epsilon =
t_{\mu\nu}u^{\mu}u^{\nu}$ is $(3/4)(\pi T)^4$. We will also
define $b$ through $b = 1/(\pi T)$. 

At equilibrium, of course, the hydrodynamic variables $u^\mu$ and $T$ are constants, and the energy-momentum tensor takes the following Lorentz-covariant
form
\begin{equation}\label{eq}
t_{\mu\nu} = \frac{3u_{\mu}u_{\nu} + P_{\mu\nu}}{4b^4},
\end{equation}  
where $P_{\mu\nu}$ is the tensor which projects to the spatial
hyperplane orthogonal to $u^{\mu}$, and is given by
\begin{equation}\label{P}
P_{\mu\nu}= u_{\mu}u_{\nu}+\eta_{\mu\nu}.
\end{equation}
It is easy to see that this energy-momentum tensor (\ref{eq}) is
traceless. The equilibrium pressure is given by $1/(4b^4)$.

The energy-momentum tensor of any non-equiibrium state can be
parametrised by
\begin{equation}
t_{\mu\nu}(z) = \frac{3u_{\mu}(z)u_{\nu}(z) +
P_{\mu\nu}(z)}{4b^4 (z)}+\pi_{\mu\nu}(z),
\end{equation}
where the hydrodynamic variables are now functions of the
boundary space and time coordinates, while $\pi_{\mu\nu}(z)$ is
the shear-stress tensor which vanishes only at equilibrium. 

At non-equilibrium, the four-velocity $u^{\mu}$ becomes the
\textit{local} velocity of energy-transport and $3/(4b^4)$
becomes the local energy-density. The shear-stress tensor
$\pi_{\mu\nu}$ should be such that it does not modify the local
energy density and the energy current, so it must satisfy
$u^{\mu}\pi_{\mu\nu} = 0$. Further the shear-stress tensor also
satisfies $Tr(\pi)=0$, because the full energy-momentum tensor
is traceless, and the equilibrium part is traceless by itself.
Therefore, the shear-stress tensor $\pi_{\mu\nu}$ actually has
only five independent components, and together with the four
hydrodynamic variables (the temperature $T$ and three
independent components of the four-velocity $u^\mu$), it can
parametrize the nine independent components of a traceless
boundary stress tensor.

To specialize to homogeneous relaxation, we make the
hydrodynamic variables constant and the shear-stress tensor
dependent only on a global time, defined through
$-u_{\mu}z^{\mu} =t$, so that
\begin{equation}\label{form2}
t_{\mu\nu}(z) = \frac{3u_{\mu}u_{\nu} +
P_{\mu\nu}}{4b^4}+\pi_{\mu\nu}(t).
\end{equation}
Given that $u^{\mu}\pi_{\mu\nu}=0$, it is easy to check that
$\partial^{\mu}t_{\mu\nu}=0$ is satisfied for any
$\pi_{\mu\nu}(t)$.

We can further simplify the above form by going to the global
comoving frame where $u^{\mu} = (1, 0, 0, 0)$. In this frame, we
have,
\begin{equation}\label{form3}
t_{00} = \frac{3}{4b^4}, \quad t_{0i} = t_{i0} = 0, \quad t_{ij}
= \frac{1}{4b^4}\delta_{ij} +\pi_{ij}(t),
\end{equation}
which is exactly the same as in (\ref{form1}) with $\epsilon =
3p = 3/(4b^4)$. In this form, it is easy to see that the
covariant $\pi_{\mu\nu}$ indeed conforms with
$\partial^{\mu}t_{\mu\nu}=0$ for any $\pi_{ij}(t)$. Also in this
frame, the traceless-ness condition becomes
$\pi_{ij}\delta_{ij}=0$. Given that the conservation of
energy and momentum is trivially satisfied and also that
the energy-momentum tensor is traceless for any traceless $\pi_{ij}(t)$, corresponding to any boundary stress tensor of the form
(\ref{form2}) or equivalently of form (\ref{form3}), a unique
bulk metric is guaranteed to exist \cite{myself1}. \footnote{It
is not so obvious given that the boundary stress-tensor is not
Cauchy data. But, it has been shown in \cite{myself1} that a
unique bulk solution exists if the bulk metric can be smoothly
connected to the AdS black brane by turning on one or more
perturbation parameters.}

The amplitude expansion parameter is the typical value of the
shear-stress tensor $\pi_{\mu\nu}$ divided by the pressure
$1/(4b^4)$. We note that the energy-momentum tensor which takes
the form (\ref{form2}), or equivalently the form (\ref{form3}),
is exactly given only by the sum of the zeroth order and the
first order terms in the amplitude expansion, where the zeroth
order term is the equilibrium part and the first order term is
the shear-stress tensor itself. Unlike the purely hydodynamic
case, the energy-momentum tensor itself does not get corrected
order by order in the expansion parameter.

At the zeroth order in the amplitude expansion, the metric is
just that of the unperturbed boosted AdS black brane, dual to a
finite temperature equilibrium state at zero chemical potentials
in the gauge theory. The metric in Fefferman-Graham coordinates
(\ref{FG}) is fully specified by $g_{\mu\nu}$. We will refer to
the AdS black brane $g_{\mu\nu}$ as $g^{(0)}_{\mu\nu}$, since it
is the zeroth order term in the amplitude expansion. It is given
by
\begin{equation}\label{zo}
g^{(0)}_{\mu\nu} = -\frac{(4b^4 -\rho^4)^2}{4b^4 (4b^4
+\rho^4)}u_{\mu}u_{\nu}+\left(1+ \frac{\rho^4}{4b^4}\right)
P_{\mu\nu}.
\end{equation}
It is easy to see from the coefficient of the $\rho^4$ term that
the boundary stress-tensor in this metric indeed takes the
equilibrium form (\ref{eq}). The coefficients of the higher
terms of the Taylor expansion about $\rho =0$ also agree with
(\ref{coeffs}), where the derivatives should be dropped.

In the global comoving frame, 
\begin{equation}\label{zof}
g^{(0)}_{00} = -\frac{(4b^4 -\rho^4)^2}{4b^4 (4b^4 +\rho^4)},
\quad g^{(0)}_{0i}=g^{(0)}_{i0} =0, \quad g^{(0)}_{ij}= \left(1+
\frac{\rho^4}{4b^4}\right) \delta_{ij}.
\end{equation}

At the first order in the amplitude expansion, the possible
corrections to $g_{\mu\nu}$ should involve tensors of rank two
which have one $\pi_{\mu\nu}$ in them. It is not hard to see that
since $u^\mu$ is constant, and $u^{\mu}\pi_{\mu\nu}=0$ by
definition of $\pi_{\mu\nu}$, we get that $u^{\mu}(u\cdot\partial)^n
\pi_{\mu\nu}$ or $u^{\mu}(d/dt)^n
\pi_{\mu\nu}$ should vanish for all $n$. Further, because
$Tr(\pi)= 0$, the corrections to the metric at the first order
in amplitude expansion can only be proportional to $\Box^n
\pi_{\mu\nu}$, or equivalently $(-)^n (d/dt)^{2n} \pi_{\mu\nu}$.
Therefore, in the global comoving frame these corrections are of
the form $(-)^n (d/dt)^{2n} \pi_{ij}$.

An easier way to arrive at the conclusion above is to go to the
global comoving frame directly, where we can readily see that
the corrections take the form $(d/dt)^{2n} \pi_{ij}$. We will stick to the
global comoving frame for the rest of this paper, until we extract the
general phenomenological equation for the shear-stress tensor in Lorentz-covariant form.
The regularity analysis of the metric is also convenient in the global comoving frame.

Further inspection of the coefficients like \eqref{coeffs} of the Taylor expansion \eqref{te} of $g_{\mu\nu}$ in the Fefferman-Graham metric \eqref{FG} shows that, at the first order in the amplitude expansion, the corrections to $g_{\mu\nu}$ should be of the form,
\begin{equation*}
\displaystyle\sum\limits_{n=0}^\infty \displaystyle\sum\limits_{m=0}^\infty b^{-m}\rho^{4+2n+m}\left(\frac{d}{dt}\right)^{2n} \pi_{ij}(t),
\end{equation*}
in the global comoving frame. Putting all our arguments together, at first order in the  amplitude expansion, $g_{\mu\nu}$ in the Fefferman-Graham metric \eqref{FG} takes the form,
\begin{equation}\label{fofs}
g^{(1)}_{00}=g^{(1)}_{ti} =
g^{(1)}_{i0} = 0, \quad g^{(1)}_{ij}=\displaystyle\sum\limits_{n=0}^\infty b^{4+2n} f^{(1, 2n)}(\rho)\left(\frac{d}{dt}\right)^{2n} \pi_{ij}(t),
\end{equation}
in the global comoving frame. 

By definition, the functions $f^{(1,2n)}(\rho)$ are dimensionless, so they should be functions of the dimensionless variable $\left(\rho/b\right)$. Also, 
\begin{equation}\label{r1}
f^{(1,2n)} = O\left(\left(\frac{\rho}{b}\right)^{4+2n}\right), \ at \ \rho=0,
\end{equation}
so that the vector constraint \eqref{ecv}  and the scalar constraint \eqref{scalar} are satisfied. Explicit calculations of the coefficients like \eqref{coeffs}, consistent with these constraints, clearly show that the leading term has to be coefficient of $(d/dt)^{2n}\pi_{ij}$, or equivalently $\Box^n t_{\mu\nu}$.

The dynamical equation \eqref{tensor} determines all the leading and subleading terms of $f^{(1,2n)}$ for all $n$, except for the leading $\rho^4$ term of $f^{(1,0)}$. The correction to the boundary stress-tensor at the first order in the amplitude expansion is $\pi_{ij}(t)$, therefore its coefficient has to be $\rho^4$. This implies
\begin{equation}\label{r2}
f^{(1,0)} = \left(\frac{\rho}{b}\right)^{4} +O\left(\left(\frac{\rho}{b}\right)^{8}\right), \ at \ \rho=0.
\end{equation}
The subleading terms of $f^{(1,0)}$  are all of $O((\rho/b)^{4n})$. This is because the coefficients like (\ref{coeffs}) of Taylor expansion of $g_{\mu\nu}$ can pick up powers of $4n$ of $\rho$ only in absence of derivatives, since $\rho^{4n}$ accompany polynomials of the boundary stress tensor. A similar logic shows that $f^{(1,2n)}$ involves powers of $4+2n+4m$ of $\rho/b$ only, for $m\geq 0$.

It will be convenient to do the Fourier transform of
$\pi_{ij}(t)$, so that,
\begin{equation}\label{Fr1}
\pi_{ij}(t) = \int_{-\infty}^{\infty}d\omega \ e^{-i\omega t} \pi_{ij}(\omega).
\end{equation}
Similarly, we can do the Fourier transforms of the components of $g^{(1)}_{\mu\nu}$. Let us further define $f^{(1)}(\rho,\omega)$ such that
\begin{equation}\label{f1}
f^{(1)}(\rho, \omega) = \displaystyle\sum\limits_{n=0}^{\infty}(-)^n b^{2n}\omega^{2n}
f^{(1,2n)}(\rho).
\end{equation}
With the above definition, $g^{(1)}_{\mu\nu}$ in Fourier space can be captured in terms of this single dimensionless function $f^{(1)}(\rho,\omega)$. This single function encodes all the time derivatives of $\pi_{ij}$, since the Fourier transform of \eqref{fofs} implies
\begin{equation}\label{fof}
g^{(1)}_{00}(\rho, \omega)=g^{(1)}_{0i}(\rho, \omega) =
g^{(1)}_{i0}(\rho, \omega) = 0, \quad g^{(1)}_{ij}(\rho,
\omega)=b^4 f^{(1)}(\rho, \omega)\pi_{ij}(\omega).
\end{equation}
The reality of $\pi_{ij}(t)$ and $g^{(1)}_{\mu\nu}$ implies that
\begin{equation}
\pi_{ij}(-\omega) = \left(\pi_{ij}(\omega)\right)^{*}, \quad f^{(1)}(\rho, -\omega)= \left( f^{(1)}(\rho, \omega)\right)^* .
\end{equation}
The latter is readily satisfied by the definition \eqref{f1} of $f^{(1)}(\rho, \omega)$, as the $f^{(1,2n)}(\rho)$'s  are real.

 The dynamical equation \eqref{tensor} determines all coefficients of Taylor expansion of
$f^{(1)}(\rho, \omega)$ given that the coefficient of the leading term $\rho^4 / b^4$ is $1$, so that the boundary stress tensor is corrected by $\pi_{ij}$. We thus find that
\begin{equation}\label{te1}
f^{(1)}(\rho, \omega) = \frac{\rho^4}{b^4}- \frac{b^2\omega^2}{12}\frac{\rho^6}{b^6} +\left(\frac{1}{4}+\frac{b^4\omega^4}{384}\right)\frac{\rho^8}{b^8} - \left(\frac{b^2\omega^2}{30}+\frac{b^6\omega^6}{23040}\right)\frac{\rho^{10}}{b^{10}}
+O\left(\rho^{12}\right).
\end{equation}
In the following subsection we will derive the equation of motion for $f^{(1)}(\rho,\omega)$, in terms of which we capture the full metric at first order in the amplitude expansion. 

Similarly, we can argue that at second order in the amplitude expansion, the corrections to $g_{\mu\nu}$ in the Fefferman-Graham metric \eqref{FG} should take the form
\begin{eqnarray}\label{sofs}
g^{(2)}_{00}&=& \displaystyle\sum\limits_{n=0}^{\infty}\displaystyle\sum\limits_{m=0}^{n} b^{8+2n}
f^{(2,2n,2m)}_3(\rho) \displaystyle\sum\limits_{\substack{a,b=0 \\ a+b=n, |a-b|=m}}^n \Bigg(\left(\frac{d}{dt}\right)^{2a}\pi_{pq}(t) \left(\frac{d}{dt}\right)^{2b}\pi_{pq}(t)\Bigg),\nonumber\\ 
g^{(2)}_{0i}&=& g^{(2)}_{i0}=0,\nonumber\\
g^{(2)}_{ij}&=&\displaystyle\sum\limits_{n=0}^{\infty}\displaystyle\sum\limits_{m=0}^{n} b^{8+2n}
f^{(2,2n,2m)}_2(\rho) \delta_{ij}\displaystyle\sum\limits_{\substack{a,b=0 \\ a+b=n, |a-b|=m}}^n \Bigg(\left(\frac{d}{dt}\right)^{2a}\pi_{pq}(t) \left(\frac{d}{dt}\right)^{2b}\pi_{pq}(t)\Bigg)\nonumber\\ 
&&+\displaystyle\sum\limits_{n=0}^{\infty}\displaystyle\sum\limits_{m=0}^{n} b^{8+2n}
f^{(2,2n,2m)}_1(\rho) \displaystyle\sum\limits_{\substack{a,b=0 \\ a+b=n, |a-b|=m}}^n \Bigg[\left(\frac{d}{dt}\right)^{2a}\pi_{ik}(t) \left(\frac{d}{dt}\right)^{2b}\pi_{kj}(t)\nonumber\\
 &&\qquad\qquad\qquad\qquad\qquad\qquad\qquad\quad \ -\frac{1}{3}\delta_{ij}\left(\frac{d}{dt}\right)^{2a}\pi_{pq}(t) \left(\frac{d}{dt}\right)^{2b}\pi_{pq}(t)\Bigg].
\end{eqnarray}
We can easily see that in the summation above over $a$ and $b$ for fixed $n$ and $m$, when $m\neq0$, $(a,b)$ is either $((n+m)/2, (n-m)/2)$, or $((n-m)/2, (n+m)/2)$, and when $m=0$, they are $(n/2, n/2)$. We see that the dimensionless functions $f^{(2,2n,2m)}_3$'s capture corrections to the time-time components of the metric, $f^{(2,2n,2m)}_2$'s capture corrections proportional to the purely spatial identity matrix, and $f^{(2,2n,2m)}_1$'s capture corrections which are purely spatial and traceless.

In Fourier space, the above can be captured in terms of just three dimensionless functions $f^{(2)}_1 (\rho,\omega,\omega_1 )$, $f^{(2)}_2 (\rho, \omega, \omega_1)$ and $f^{(2)}_3 (\rho, \omega, \omega_1)$ as 
\begin{eqnarray}\label{sof}
g^{(2)}_{00}&=& b^8 \int_{-\infty}^{\infty}d\omega \ e^{-i\omega t}\int_{-\infty}^{\infty}d\omega_1 
\ f^{(2)}_3(\rho,\omega, \omega_1) \pi_{pq}(\omega_1 )\pi_{pq}(\omega-\omega_1 ), \nonumber\\
g^{(2)}_{0i}&=& g^{(2)}_{i0}=0,\nonumber\\
g^{(2)}_{ij}&=&b^8\delta_{ij} \int_{-\infty}^{\infty}d\omega \ e^{-i\omega t}\int_{-\infty}^{\infty}d\omega_1 
\ f^{(2)}_2(\rho,\omega, \omega_1) \pi_{pq}(\omega_1 )\pi_{pq}(\omega-\omega_1 )\nonumber\\
&&+\frac{b^8}{2} \int_{-\infty}^{\infty}d\omega \ e^{-i\omega t}\int_{-\infty}^{\infty}d\omega_1 \ 
f^{(2)}_1(\rho,\omega, \omega_1)\Big[ \pi_{ik}(\omega_1 ) \ \pi_{kj}(\omega-\omega_1 )\nonumber\\
&&\ \ \ \ \ \ \ \ \ \ \ \ \ \ \ \ \ \ \ \ \ \ \ \ \ \ \ \ \ \ \ \ \ \ \ \ +\pi_{ik}(\omega-\omega_1 )\pi_{jk}(\omega_1 ) \nonumber\\
&&\ \ \ \ \ \ \ \ \ \ \ \ \ \ \ \ \ \ \ \ \ \ \ \ \ \ \ \ \ \ \ \ \ \ \ \ -\frac{2}{3}\delta_{ij}\pi_{rs}(\omega_1 )\pi_{rs}(\omega-\omega_1 )\Big],
\end{eqnarray}
where
\begin{eqnarray}\label{f2i}
f^{(2)}_i (\rho,\omega,\omega_1) &=& \displaystyle\sum\limits_{n=0}^{\infty}\displaystyle\sum\limits_{m=0}^{n}(-)^n b^{2n}\left(\omega_1^{n+m}(\omega-\omega_1)^{n-m}+\omega_1^{n-m}(\omega-\omega_1)^{n+m}\right)\nonumber\\
&&f^{(2,2n,2m)}_i (\rho), \qquad \text{for} \ i=1,2,3.
\end{eqnarray}
It is important to note that the definitions of $f^{(2)}_i (\rho, \omega, \omega_1)$ are such
that they is symmetric under the exchange of $\omega_1$ and $\omega - \omega_1$. The powers of $\omega_1$ above denote time derivatives acting on the first $\pi$ and the powers of $\omega-\omega_1$ denote time derivatives acting on the second $\pi$, so
it is natural to define these functions in the symmetric fashion mentioned.

In the following subsections we will obtain equations of motion for these $f^{(2)}_i (\rho,\omega, \omega_1)$'s. However, from the discussion so far, it is easy to see that the dynamical equation \eqref{tensor} determines all the coefficients of $f^{(2)}_i (\rho,\omega, \omega_1)$'s  in the Taylor expansion about $\rho = 0$, given the Taylor series for $f^{(1)}(\rho,\omega)$ has been obtained previously at the first order. We thus obtain,
\begin{eqnarray}\label{te2}
f^{(2)}_{1}(\rho, \omega,\omega_1) &=& \frac{\rho^8}{2b^8}
-\frac{b^2(\omega_1^{2} + (\omega
-\omega_1)^2)}{24}\frac{\rho^{10}}{b^{10}}+O\left(\rho^{12}\right)
,\nonumber\\
f^{(2)}_{2}(\rho, \omega,\omega_1) &=& \frac{\rho^8}{8b^8}
-\left(\frac{b^2(\omega_1^{2} + (\omega
-\omega_1)^2)}{90}-\frac{b^2\omega_1 (\omega
-\omega_1)}{720}\right)\frac{\rho^{10}}{b^{10}}+O\left(\rho^{12}\right),
\nonumber\\
f^{(2)}_{3}(\rho, \omega,\omega_1) &=& \frac{\rho^8}{24b^8}
-\left(\frac{b^2(\omega_1^{2} + (\omega
-\omega_1)^2)}{240}-\frac{b^2\omega_1 (\omega
-\omega_1)}{240}\right)\frac{\rho^{10}}{b^{10}}+O\left(\rho^{12}\right).
\end{eqnarray}
We can also check that the vector constraint \eqref{ecv} and the scalar constraint \eqref{scalar} vanish at the second order in the amplitude expansion, if we substitute the above expansions in them and Taylor expand about $\rho=0$. We also note that the above expansions are $O(\rho^8)$ at the leading order, so that the boundary stress-tensor receives no corrections at this order as discussed before.

We note that the reality of $g^{(2)}_{\mu\nu}$ implies
\begin{equation}
f^{(2)}_i (\rho, -\omega, -\omega_1) = \left(f^{(2)}_i (\rho, \omega, \omega_1)\right)^* .
\end{equation}
The above is guaranteed by the definitions \eqref{f2i}.

We can similarly obtain the forms of the expansion of $g_{\mu\nu}$ at higher orders in the amplitude expansion. The corrections can be always grouped into three categories, namely the corrections in the time-time component, the corrections which are proportional to the spatial identity matrix, and finally the corrections which are purely spatial and traceless. In each category, we can have one or more independent tensor
structures. However, after Fourier transforming the dependence on the time coordinate, all these corrections can be captured by a finitely few functions, which also efficiently sum over all time derivatives.

\subsection{The equations of motion and their solutions in the
amplitude expansion}

We proceed to obtain the explicit equations of motion of the metric, order by order in the amplitude expansion, summing over all time derivatives. We will see that in the Fefferman-Graham coordinates, one can generate a simple algorithm to satisfy the constraints.

In the previous section we made a distinction between the true dynamical equation \eqref{tensor}, and the four-dimensional tensorial components \eqref{ed} of Einstein's equation. The former was obtained by suitably combining the latter with the scalar component \eqref{ecs} of Einstein's equation multiplied by the four-dimensional metric $g$. This distinction will not matter much in practice. The tensor equation \eqref{ed} is almost as good as the true dynamical equation \eqref{tensor} in determining the coefficients $g_{(2n)\mu\nu}$ of the Taylor expansion \eqref{te} of $g_{\mu\nu}$ for $n>2$, given a flat boundary metric $g_{(0)\mu\nu}$ and a particular boundary stress-tensor which satisfies energy-momentum conservation, is traceless, and is identified with $g_{(4)\mu\nu}$. It is almost as good because it fails only in the case of one coefficient of the Taylor expansion, for $n>2$, which is $g_{(8)\mu\nu}$. It will turn out that this will 
just introduce a slight modification in the general scheme, only at the second order in the amplitude expansion. 

We will use the tensor equation \eqref{tensor} therefore to obtain the equations of motion, order by order in the amplitude expansion Once we impose the correct boundary conditions for the equations of motion, the solutions will automatically satisfy the vector constraint \eqref{ecv} and the scalar constraint \eqref{scalar}. Also, we will not make a distinction between checking the true scalar constraint \eqref{scalar} and the scalar component of Einstein's equation \eqref{ecs}. The former is obtained by combining the latter with the trace of the tensor equation \eqref{ed}, which we are now treating as the dynamical equation. So, it is sufficient to check if the latter is satisfied.  

\subsubsection{The metric at the first order in the amplitude expansion}
At the first order in the amplitude expansion, we recall that $g_{\mu\nu}$ takes the
form \eqref{fofs}. After Fourier transforming the dependence on the time coordinate,
the metric takes the compact form \eqref{fof}, which involves a single function $f^{(1)}(\rho, \omega)$ defined in \eqref{f1}. We can readily find the equation of motion of 
$f^{(1)}(\rho, \omega)$ by expanding the tensor equation up to first order in the amplitude expansion and then Fourier transforming the time dependence. 

Let us define the differential operator $D_{1\omega}$ through
\begin{equation}\label{d1}
D_{1\omega} = \frac{\partial^2}{\partial \rho^2} + \La \frac{\partial}{\partial \rho} + \Lb + \omega^2 \Lc,
\end{equation} 
where,
\begin{eqnarray}\label{l1}
\La &=& - \frac{(12b^4 - \rho^4)(4b^4 + 3 \rho^4)}{\rho(4b^4 - \rho^4)(4b^4 + \rho^4)}, \quad \Lb = \frac{128 \rho^6 b^4}{(4b^4 - \rho^4)(4b^4 + \rho^4)^2}, \nonumber\\
\Lc &=& \frac{4b^4(4b^4 + \rho^4)}{(4b^4 - \rho^4)^2}.
\end{eqnarray}
The equation of motion of $f^{(1)}(\rho, \omega)$ obtained from the Fourier transform of the tensor equation is \eqref{ed} is
\begin{equation}\label{eom1}
D_{1\omega}f^{(1)}(\rho, \omega) = 0.
\end{equation}

At the first order in the amplitude expansion, the vector equation \eqref{ecv} and the scalar equation \eqref{ecs} identically vanish. It follows from the definitions of $\pi_{ij}$, that in Fourier space $\pi_{ij}(\omega)\delta_{ij}= 0$. A simple inspection reveals this makes the both the vector and scalar equations vanish. 

However, in order that the solution can be extended at higher orders in the amplitude expansion, without conflicting with the constraints, and also because the correction to the boundary stress-tensor at first order in the amplitude expansion has to be $\pi_{ij}$, we need the solution to the equation of motion \eqref{eom1} for $f^{(1)}(\rho, \omega)$ to 
satisfy the boundary condition 
\begin{equation}\label{bc1}
f^{(1)}(\rho, \omega) =  \left(\frac{\rho}{b}\right)^{4} +O\left(\left(\frac{\rho}{b}\right)^{6}\right), \text{at} \ \rho=0.
\end{equation}
One can readily see that the above boundary condition specifies the solution of $f^{(1)}(\rho,\omega)$ uniquely, by Taylor expanding eq. \eqref{eom1} in $\rho$ about $\rho=0$ and checking this determines all the coefficients of $\rho^{2n}$ of the Taylor expansion of $f^{(1)}(\rho,\omega)$ for $n>3$. One can also check that this reproduces the known Taylor series of \eqref{te1} of $f^{(1)}(\rho, \omega)$.

To sum up, we have thus determined that at the first order the metric in Fourier space is given by $f^{(1)}(\rho, \omega)$, which is the unique solution of eq. \eqref{eom1} with the boundary condition \eqref{bc1}. This solution sums up the time derivatives to all orders.

Unfortunately, \eqref{eom1} is not exactly solvable unless $\omega=0$. We can, however, devise the following strategy. We use \eqref{f1} which relates $f^{(1)}(\rho, \omega)$ to the
metric in real time. This relation actually is also the Taylor expansion of $f^{(1)}(\rho,\omega)$ in $\omega$ about $\omega=0$. It is easy to see that \eqref{eom1} in conjunction with \eqref{f1} implies that
\begin{equation}\label{rec1}
\hat{D}_1 f^{(1, 0)} = 0,
\end{equation}
and
\begin{equation}\label{rec2}
\hat{D}_1 f^{(1, 2n)} = \frac {\Lc}{b^2}   f^{(1, 2(n-1))}, \text{for} \ n\geq1,
\end{equation}
where $\hat{D}$ is the differential operator $D_{1\omega}$ at $\omega =0$, i.e.
\begin{equation}\label{d1hat}
\hat{D}_1 =\frac{d^2}{d \rho^2} + \La \frac{d}{d \rho} + \Lb,
\end{equation}
with $L_{11}$ and $L_{12}$ as defined in \eqref{l1}. Also, the boundary condition \eqref{bc1} implies \eqref{r1} and \eqref{r2}.

The solutions of the homogeneous eq. \eqref{rec1} are
\begin{equation}\label{s}
s_1 (\rho) = 1 + \frac{\rho^4}{4b^4}, \quad s_2 (\rho) = \left(1 + \frac{\rho^4}{4b^4}\right)
\log \left(\frac{4b^4 - \rho^4}{4b^4 + \rho^4}\right).
\end{equation}
The boundary condition \eqref{r2} for $f^{(1,0)}$ implies that
\begin{equation}\label{sol1}
f^{(1, 0)}(\rho) = -2 s_2 (\rho) = -2\left(1 + \frac{\rho^4}{4b^4}\right)
\log \left(\frac{4b^4 - \rho^4}{4b^4 + \rho^4}\right).
\end{equation}
The eqs. \eqref{rec2} alongwith the boundary conditions \eqref{r1} imply the following recursion series
\begin{eqnarray}\label{soln}
 f^{(1, 2n)}(\rho) &=& \frac{1}{b^2} (-s_1 (\rho) \int_{0}^{\rho} d\rho^{'} \frac{s_2 (\rho^{'})}{W(s_1, s_2)(\rho^{'})} L_{13} (\rho^{'}) f^{(1, 2(n-1))}(\rho^{'})\nonumber\\
&&+ s_2 (\rho) \int_{0}^{\rho} d\rho^{'} \frac{s_1 (\rho^{'})}{W(s_1, s_2)(\rho^{'})} L_{13} (\rho^{'}) f^{(1, 2(n-1))}(\rho^{'}) ), \ \text{for} \ n\geq1,
\end{eqnarray}
where $W(s_1 , s_2)$ is the Wronskian of $s_1$ and $s_2$.

Through the exact recursion relations given by \eqref{sol1} and \eqref{soln}, we thus
efficiently sum over all time derivatives in the metric at first order in the amplitude
expansion, or equivalently define the coefficients of the Taylor series \eqref{f1}
of $f^{(1)}(\rho, \omega)$  about $\omega=0$.

\subsubsection{The metric at the second order in the amplitude expansion}

At the second order in the amplitude expansion, we have seen that we can express $g_{\mu\nu}$ in the Fefferman-Graham metric \eqref{FG} in terms of three functions 
$f^{(2)}_i (\rho, \omega, \omega_1 )$, for $i=1,2,3$ after Fourier transforming the time dependence as shown earlier in \eqref{sof}.

The tensor equation \eqref{ed} gives the equation of motion for all three $f^{(2)}_i$'s.  When 
we Fourier transform the dependence on time of this equation, to be consistent with
the definitions of $f^{(2)}_i$'s, the Fourier transform is done such that it is symmetric under
the exchange of $\omega_1$ and $\omega-\omega_1$.

The part of this equation of motion which is purely spatial and traceless at the second order in the amplitude expansion, gives the equation of motion for $f^{(2)}_1$,
\begin{equation}\label{eom21}
D_{1\omega} f_{1}^{(2)} \left(\rho, \omega,\omega_{1}\right) =  S_{1}^{(2)}\left(\rho, \omega,\omega_{1}\right),
\end{equation}
where $D_{1\omega}$ is defined through \eqref{d1}, and
\begin{eqnarray}
S_{1}^{(2)}\left(\rho, \omega,\omega_{1}\right) &=& \Rb \left(\frac{\partial}{\partial \rho} f^{(1)} (\rho, \omega_1) \frac{\partial}{\partial \rho} f^{(1)} (\rho, \omega - \omega_1) \right) \nonumber\\
&&+ \Rc \left(f^{(1)} (\rho,\omega_1) \frac{\partial}{\partial \rho} f^{(1)} (\rho,\omega - \omega_1) + f^{(1)} (\rho,\omega - \omega_1)\frac{\partial}{\partial \rho}f^{(1)} (\rho,\omega_1)  \right) \nonumber\\
&& + \left( \Rd + \omega_1(\omega - \omega_1) \Re\right)  f^{(1)} (\rho,\omega_1) f^{(1)} (\rho, \omega - \omega_1) , 
\end{eqnarray}
with
\begin{eqnarray}\label{rfnseom1}
\Rb &=& \frac{4b^4}{(4b^4 + \rho^4)}, \quad \Rc = - \frac{16 \rho^3 b^4}{(4b^4 + \rho^4)^2} ,\nonumber\\
 \Rd &=& \frac{64 \rho^6 b^4}{(4b^4 + \rho^4)^3}, \quad \Re = \frac{16b^8}{(4b^4 - \rho^4)^2}  . 
\end{eqnarray} 

Since $f^{(2)}_1$ captures the contributions to the metric which are purely spatial
and traceless (just like the entire contribution of the metric at the first order in the amplitude expansion), it follows that its equation of motion should be given by the same 
differential operator $D_{1\omega}$ acting on it, but with source terms on the right hand
side. In fact, this structure of the equation of motion for the purely spatial and traceless parts is maintained to all orders in the amplitude expansion.

The equations of motion  couple $f^{(2)}_2$ and $f^{(2)}_3$. Let us define the following differential operators
\begin{eqnarray}
D_{21\omega} &=& \frac{\partial^2}{\partial \rho^2} + \LLa \frac{\partial}{\partial \rho} + \LLb + \omega^2 \LLc , \nonumber\\
D_{22} &=& \LLd \frac{\partial}{\partial \rho} + \LLe, \nonumber\\
D_{31}&=& \frac{\partial^2}{\partial \rho^2} + \Laa \frac{\partial}{\partial \rho} + \Lbb , \nonumber\\
D_{32\omega} &=& \Lcc \frac{\partial}{\partial \rho} + \Ldd + \omega^2 \Lee ,
\end{eqnarray}
with
\begin{eqnarray}\label{l2}
\LLa &=& -\frac{8b^4(12b^4 + \rho^4)}{\rho(4b^4 - \rho^4)(4b^4 + \rho^4)}, \quad \LLb = \frac{4\rho^2(48b^8 + 8 \rho^4 b^4 + 3 \rho^8)}{(4b^4 - \rho^4)(4b^4 + \rho^4)^2} , \nonumber\\
\LLc &=& \frac{4b^4(4b^4 + \rho^4)}{(4b^4 - \rho^4)^2}, \quad \LLd = \frac{(4b^4 + \rho^4)}{\rho(4b^4 - \rho^4)}, \quad \LLe = \frac{4\rho^2 (12b^4 + \rho^4)}{(4b^4 - \rho^4)^2} , \nonumber\\
\Laa &=& - \frac{2(32b^8 - 36 \rho^4 b^4 - \rho^8)}{\rho(4b^4 - \rho^4)(4b^4 + \rho^4)}, \quad \Lbb = -\frac{4 \rho^2 (192 b^{12} - 336 \rho^4 b^8 - 60 \rho^8 b^4 + \rho^{12})}{(4b^4 - \rho^4)^2(4b^4 + \rho^4)^2} , \nonumber\\
\Lcc &=& \frac{3(4b^4 - \rho^4)(16 b^8 + 24 \rho^4 b^4 + \rho^8)}{\rho (4b^4 + \rho^4)^3} , \nonumber\\ \Ldd &=& - \frac{12 \rho^2 (4b^4 - \rho^4)(16b^8 + 24 \rho^4 b^4 + \rho^8)}{(4b^4 + \rho^4)^4} , \quad
\Lee = -\frac{12b^4}{(4b^4 + \rho^4)}.
\end{eqnarray}

The part of the tensor equation \eqref{ed} which is proportional to the purely spatial identity matrix gives
\begin{equation}\label{eom22}
D_{21\omega} f_{2}^{(2)} \left(\rho, \omega,\omega_{1}\right) + D_{22}f_{3}^{(2)}\left(\rho, \omega,\omega_{1}\right)  =  S_{2}^{(2)}\left(\rho, \omega,\omega_{1}\right),
\end{equation}
 where
\begin{eqnarray}
S_{2}^{(2)}\left(\rho, \omega,\omega_{1}\right) &=& \RRb \left(\frac{\partial}{\partial \rho} f^{(1)} (\rho,\omega_1) \frac{\partial}{\partial \rho} f^{(1)} (\rho,\omega - \omega_1) \right) \nonumber\\
&&+ \RRc \left(f^{(1)} (\rho,\omega_1) \frac{\partial}{\partial \rho} f^{(1)} (\rho,\omega - \omega_1) + f^{(1)} (\rho, \omega - \omega_1) \frac{\partial}{\partial \rho} f^{(1)} (\rho,\omega_1) \right) \nonumber\\
&& + \left( \RRd + \omega_1(\omega - \omega_1) \RRe\right)  f^{(1)} (\rho,\omega_1) f^{(1)} (\rho, \omega - \omega_1),
\end{eqnarray}
with
\begin{eqnarray}\label{rfnseom2}
\RRb &=& \frac{4b^4}{3(4b^4 + \rho^4)},\quad \RRc = -\frac{2b^4(12b^4 + 5\rho^4)}{3\rho(4b^4 + \rho^4)^2} , \nonumber\\
\RRd &=& \frac{16 \rho^2 b^4 (12b^4 + \rho^4)}{3 (4b^4 + \rho^4)^3}, \quad \RRe = \frac{16 b^8}{3(4b^4 - \rho^4)^2} .
\end{eqnarray}

The time-time component of the tensor equation \eqref{ed} gives
\begin{equation}\label{eom23}
D_{31} f_{3}^{(2)} \left(\rho, \omega,\omega_{1}\right) + D_{32\omega} f_{2}^{(2)}\left(\rho, \omega,\omega_{1}\right)   =  S_{3}^{(2)}\left(\rho, \omega,\omega_{1}\right),
\end{equation}
where
\begin{eqnarray}
S_{3}^{(2)}\left(\rho, \omega,\omega_{1}\right)&=&  \Rbb \left(f^{(1)} (\rho, \omega_1) \frac{\partial}{\partial \rho} f^{(1)} (\rho, \omega - \omega_1) + f^{(1)} (\rho, \omega - \omega_1) \frac{\partial}{\partial \rho} f^{(1)} (\rho, \omega_1) \right) \nonumber\\
&& + \left( \Rcc +\left( \omega_1^{2} + (\omega - \omega_1)^2 + \omega_1(\omega- \omega_1)\right) \Rdd\right) \nonumber\\
&& f^{(1)} (\rho,\omega_1) f^{(1)} (\rho, \omega - \omega_1),
\end{eqnarray}
with
\begin{eqnarray}\label{rfnseom3}
\Rbb &=& \frac{2b^4 (4b^4 - \rho^4)(16b^8 + 24 \rho^4 b^4 + \rho^8)}{\rho(4b^4 + \rho^4)^4}, \nonumber\\
\Rcc &=& -\frac{16 \rho^2b^4(4b^4 - \rho^4)(16 b^8 + 24 \rho^4 b^4 + \rho^8)}{(4b^4 + \rho^4)^5}, \quad \Rdd = - \frac{8b^8}{(4b^4 + \rho^4)^2} .
\end{eqnarray}
The eqs. \eqref{eom21}, \eqref{eom22}, and \eqref{eom23} sum the dynamical equations of motion.

As discussed previously, to be consistent with the constraints, we require that
\begin{equation}\label{bc21}
f^{(2)}_1 (\rho, \omega, \omega_1 ) = O\left(\left(\frac{\rho}{b}\right)^8\right), \text{at} \ \rho=0.
\end{equation}
The equation of motion \eqref{eom21} determines the leading and the subleading terms of the expansion. We require the same asymptotic behaviour of $f^{(2)}_2$ and $f^{(2)}_3$, however the eqs. \eqref{eom22} and \eqref{eom23} only determine the sum of the
coefficients of $\rho^8$ of $f^{(2)}_2$ and $f^{(2)}_3$. We have already discussed that the
origin of this incapability is that the parent tensor eq. \eqref{ed} from which these are  derived is not the true dynamical equation given by \eqref{tensor}. 

This incapability is easily rectifiable, since the tensor eq. \eqref{ed} is only incapable of determining the full coefficient of the $\rho^8$ term in the Taylor expansion of $g_{\mu\nu}$ about $\rho = 0$, in terms of the boundary metric
and boundary stress tensor. We should therefore use the true dynamical equation \eqref{tensor} to determine the coefficient of $\rho^8$ in Taylor series of either
$f^{(2)}_2$ or $f^{(2)}_3$, then the eqs. \eqref{eom22} and \eqref{eom23} 
determine the other coefficient and all other subleading terms of both the expansions.
So, the boundary conditions of $f^{(2)}_2$ and $f^{(2)}_3$ are
\begin{eqnarray}\label{bc22}
f^{(2)}_2 (\rho, \omega, \omega_1 ) &=& \frac{\rho^8}{8b^8}+O\left(\left(\frac{\rho}{b}\right)^{10}\right), \text{at} \ \rho=0, \nonumber\\
f^{(2)}_3 (\rho, \omega, \omega_1 ) &=& O\left(\left(\frac{\rho}{b}\right)^8\right), \text{at} \ \rho=0 .
\end{eqnarray}
One can check that, with the boundary conditions \eqref{bc21} and \eqref{bc22}, the equations of motion \eqref{eom21}, \eqref{eom22} and \eqref{eom23} reproduce the known Taylor series \eqref{te2} of $f^{(2)}_i$'s that are consistent with the constraints. Also, these 
boundary conditions make it explicit that the boundary stress-tensor is uncorrected at this 
order.

We have also observed before that checking the constraints is equivalent to checking the vector eq. \eqref{ecv} and the scalar eq. \eqref{ecs}. We have reproduced these equations
at the second order in the amplitude expansion in appendix A. These equations vanish if we substitute the solutions of the dynamical eqs. \eqref{eom21}, \eqref{eom22} and \eqref{eom23}, determined uniquely by the boundary conditions \eqref{bc21} and \eqref{bc22} in them. We can readily check this by Taylor expanding these equations in
$\rho$ about $\rho=0$ and substituting the Taylor expansions \eqref{te2} of $f^{(2)}_i$'s in them. 
 
To sum up, we have captured $g_{\mu\nu}$ in the Fefferman-Graham metric \eqref{FG}
at the second order in the amplitude expansion \eqref{sof}, through three functions 
$f^{(2)}_i (\rho, \omega, \omega_1)$, for $i=1,2,3$, which are determined uniquely
by the equations of motion \eqref{eom21}, \eqref{eom22} and \eqref{eom23}, along with
the boundary conditions \eqref{bc21} and \eqref{bc22}. This allows us to sum over all
time derivatives at this order. We also note that all of these equations along with the constraints, after Fourier transforming the time dependence, are local in both the arguments $\omega_1$ and $\omega_1$, simply because these equations are local in time as well.

Unfortunately again, this system of linear ODEs comprising \eqref{eom21}, \eqref{eom22} and \eqref{eom23}, is not solvable exactly unless $\omega=\omega_1 = 0$. Therefore,
we repeat our strategy at the first order to sum over the $\omega$ and $\omega_1$ dependence through an exact recursion series.

We use the previously defined Taylor expansion of the $f^{(2)}_i(\rho, \omega, \omega_1)$'s in $\omega$ and $\omega_1$ as defined in \eqref{f2i}, and use the full equations of motion \eqref{eom21}, \eqref{eom22} and \eqref{eom23} to find equations of motion for the coefficients $f^{(2, 2n, 2m)}_i (\rho)$. We recall that, by definition $0\leq m \leq n$.

The first equation of motion \eqref{eom21} along with \eqref{f2i} implies
\begin{eqnarray}\label{e11}
\hat{D}_{1} f_{1}^{(2, 0,0)} &=& \Rb \left(\d f^{(1,0)}\right)^2 + 2\Rc\left( f^{(1,0)} \d f^{(1,0)}\right) \nonumber\\ &&+ \Rd \left(f^{(1,0)}\right)^2,
\end{eqnarray}
when $n = m = 0$, $\hat{D}_1$ is as defined in \eqref{d1hat}, 
\begin{eqnarray}\label{e12}
\hat{D}_{1} f_{1}^{(2,2n,2m)} &=&  \frac{\Lc}{b^2} \Big(f_{1}^{(2,2n-2, 2m-2)} + 2 f_{1}^{(2,2n-2, 2m)} \Theta(n-m-1)\nonumber\\ &&\qquad\quad\quad+ f_{1}^{(2,2n-2, 2m+2)} \Theta(n-m-2) \Big) \nonumber\\&&+ \Rb \d f^{(1,n+m)} \d f^{(1,n-m)}\nonumber\\
&&+ \Rc\left( f^{(1,n+m)} \d f^{(1,n-m)} + f^{(1,n-m)} \d f^{(1,n+m)}\right) \nonumber\\
&&+\Rd f^{(1,n-m)} f^{(1,n+m)} , \ n\geq1, \ \text{(m + n) = 0(mod 2)},
\end{eqnarray}
and
\begin{eqnarray}\label{e13}
\hat{D}_{1}f_{1}^{(2,2n,2m)} &=&  \frac{\Lc}{b^2} \Big(f_{1}^{(2,2n-2, 2m-2)} + 2 f_{1}^{(2,2n-2, 2m)} \Theta(n-m-1)\nonumber\\ &&\qquad\quad\quad+ f_{1}^{(2,2n-2, 2m+2)} \Theta(n-m-2) \Big) \nonumber\\
&&-\frac{ \Re}{b^2} f^{(1,n-m-1)} f^{(1,n+m-1)} ,  \ n\geq1,  \text{(m + n) = 1(mod 2)}.
\end{eqnarray} 
The terms with $\Theta(n-m-1)$ contribute only when $n\geq m+1$ and those with $\Theta(n-m-2)$ contribute only when $n\geq m+2$. The $R_{1i}$'s above have been defined in \eqref{rfnseom1} and $L_{13}$ has been defined in \eqref{l1}. Also the $f^{(1,p)}$'s are as obtained in \eqref{sol1} and \eqref{soln} at the first order, and contribute only when $p$ is even and non-negative, otherwise they vanish.

The boundary condition \eqref{bc21} implies the solutions
\begin{eqnarray}\label{sol21}
 f^{(1, 2n, 2m)}_1 (\rho) &=& -s_1 (\rho) \int_{0}^{\rho} d\rho^{'} \frac{s_2 (\rho^{'})}{W(s_1, s_2)(\rho^{'})} S^{(2,2n, 2m)}_1 (\rho^{'})\nonumber\\
&&+ s_2 (\rho) \int_{0}^{\rho} d\rho^{'} \frac{s_1 (\rho^{'})}{W(s_1, s_2)(\rho^{'})}S^{(2,2n,2m)}_1 (\rho^{'}),
\end{eqnarray}
where $s_1$ and $s_2$ are the solutions of the homogeneous eq $\hat{D}_1 f =0$ as defined in \eqref{s}, $W(s_1, s_2)$ is their Wronskian, and $S^{(2, 2n, 2m)}_1$ is whatever
that appears in the right hand side of equations of motion \eqref{e11}, \eqref{e12} and \eqref{e13} for $f^{(2, 2n, 2m)}_1$'s, according to whether $n=m=0$, or $n\geq1$ and $n+m=0(\mod \ 2)$, or  $n\geq1$ and $n+m=1(\mod \ 2)$, respectively. When $n=m=0$, the integration above can be explicitly done, so that
\begin{equation}
f^{(2,0,0)}_1 (\rho) = 2\left(1 + \frac{\rho^4}{4b^4}\right)
\ \left( Log \left(\frac{4b^4 - \rho^4}{4b^4 + \rho^4}\right)\right)^2.
\end{equation}

To find the equations of motion for $f^{(2, 2n, 2m)}_2$'s and $f^{(2,2n, 2m)}_3$'s, it is
convenient to first decouple the equations of motion \eqref{eom22} and \eqref{eom23} for 
$\omega=\omega_1 = 0$ and then use the expansions \eqref{f2i}. 

Let us define the differential operator $\hat{D}_2$ through
\begin{equation}\label{dhat2}
\hat{D}_2 = \frac{d^3}{d \rho^3} - \frac{(144b^8+ 16 \rho^4 b^4 + 3\rho^8)}{\rho(4b^4 + \rho^4)(4b^4 - \rho^4)} \frac{d^2}{d \rho^2} + \frac{3(112b^8 + 16 \rho^4 b^4 + 5 \rho^8)}{\rho^2(4b^4 + \rho^4)(4b^4 - \rho^4)} \frac{d}{d \rho},
\end{equation}
  and the differential operator $\hat{D}_3$ through
\begin{equation}\label{dhat3}
\hat{D}_3 = \frac{d}{d \rho} + \frac{4 \rho^3 (12 b^4 + \rho^4)}{ (4b^4 + \rho^4)(4b^4 - \rho^4)}.
\end{equation}
The solutions of the homogeneous differential equation $\hat{D}_2 f = 0$ are
\begin{equation}\label{t}
t_1 = 1, \quad t_2 =1 + \frac{\rho^4}{4b^4}, \quad t_3 = \left(1 + \frac{\rho^4}{4b^4}\right)
\log \left(\frac{4b^4 - \rho^4}{4b^4 + \rho^4}\right) .
\end{equation}
The solution of the homogeneous differential equation $\hat{D}_3 f = 0$ is 
\begin{equation}\label{u}
u = \frac{(4b^4-\rho^4)^2}{4b^4(4b^4 +\rho^4)}\ .
\end{equation}

The equations of motion for $f^{(2, 2n, 2m)}_2$'s and $f^{(2, 2n, 2m)}_3$ can be written as
\begin{eqnarray}\label{e14}
\hat{D}_2 f^{(2, 2n, 2m)}_2 &=&  S^{(2, 2n, 2m)}_2 ,\nonumber\\
\hat{D}_3 f_{3}^{(2,2n,2m)} &=& S^{(2, 2n, 2m)}_3 ,
\end{eqnarray}
where
\begin{eqnarray}\label{sc}
S^{(2, 2n, 2m)}_2 &=& \left(\frac{d}{d \rho} - \frac{(48b^8 + 8 \rho^4 b^4 + 3 \rho^8)}{\rho (4b^4 + \rho^4)(4b^4 - \rho^4)}\right)T_{2}^{(2)} (\rho) - \LLd T_3^{(2)}(\rho), \nonumber\\
S^{(2, 2n, 2m)}_3 &=& \frac{1}{\LLd}\left(T_2^{(2)}(\rho) - \left(\frac{d^2}{d \rho^2} + \LLa \frac{d}{d \rho} + \LLb \right) f_{2}^{(2,2n,2m)}\right),
\end{eqnarray}
with
\begin{eqnarray}
T_2^{(2)}(\rho)  &=&  \RRb \left(\d f^{(1,0)} \right)^2 + 2 \RRc \left( f^{(1,0)} \d f^{(1,0)}\right) \nonumber\\ &&+ \RRd\left( f^{(1,0)}\right)^2, \ \text{when} \ n=m=0,
\nonumber\\
 &=&  \frac{\LLc}{b^2}\Big(f_{2}^{(2,2n-2, 2m-2)} + 2 f_{2}^{(2,2n-2, 2m)} \Theta(n-m-1)\nonumber\\ &&\qquad\quad\quad+ f_{2}^{(2,2n-2, 2m+2)} \Theta(n-m-2) \Big)  \nonumber\\ &&+ \RRb \d f^{(1,n-m)} \d f^{(1,n+m)} \nonumber\\
&&+ \RRc \left( f^{(1,n+m)} \d f^{(1,n-m)} + f^{(1,n-m)} \d f^{(1,n+m)}\right)\nonumber \\ 
&&+ \RRd f^{(1,n-m)} f^{(1,n+m)} ,\ \text{when} \ n\geq1, \ \text{(m+n) = 0(mod 2)} \nonumber\\
 &=&  \frac{\LLc}{b^2} \Big(f_{2}^{(2,2n-2, 2m-2)} + 2 f_{2}^{(2,2n-2, 2m)} \Theta(n-m-1)\nonumber\\ &&\qquad\quad\quad+ f_{2}^{(2,2n-2, 2m+2)} \Theta(n-m-2)\Big)   \nonumber\\
&&- \frac{\RRe}{b^2} f^{(1,n-m-1)} f^{(1,n+m-1)}, \text{when} \ n\geq1, \ \text {(m+n) = 1(mod 2)},
\end{eqnarray}
and
\begin{eqnarray}
T_3^{(2)}(\rho) &=& 2 \Rbb  \left( f^{(1,0)} \d f^{(1,0)} \right)\nonumber\\
&& + \Rcc \left(f^{(1,0)}\right)^2, \ \text{when} \ n=m=0, \nonumber\\ 
&=&  \frac{\Lee}{b^2}  \Big(f_{2}^{(2,2n-2, 2m-2)} + 2 f_{2}^{(2,2n-2, 2m)} \Theta(n-m-1)\nonumber\\ &&\qquad\quad\quad+ f_{2}^{(2,2n-2, 2m+2)} \Theta(n-m-2) \Big)  \nonumber\\
&&+ \Rbb  \left( f^{(1,n+m)} \d f^{(1,n-m)} + f^{(1,n-m)} \d f^{(1,n+m)}\right) \nonumber\\
&& + \Rcc f^{(1,n-m)} f^{(1,n+m)} \nonumber\\ 
&&- \frac{\Rdd}{b^2} \Big(f^{(1,n-m-2)} f^{(1,n+m)} \nonumber\\&&\qquad\quad\quad+ f^{(1,n-m)} f^{(1,n+m-2)}\Big),\ \text{when} \ n\geq1, \ \text{(m+n) = 0(mod 2)},\nonumber\\
 &=& \frac{\Lee}{b^2}  \Big(f_{2}^{(2,2n-2, 2m-2)} + 2 f_{2}^{(2,2n-2, 2m)} \Theta(n-m-1)\nonumber\\ &&\qquad\quad\quad+ f_{2}^{(2,2n-2, 2m+2)} \Theta(n-m-2)\Big) \nonumber\\&&-\frac{ \Rdd}{b^2} f^{(1,n-m-1)} f^{(1,n+m-1)}, \text{when} \ n\geq1, \ \text{(m+n) = 1(mod 2)}.
\end{eqnarray}
In the above eqs. $R^{(2)}_{2i}$'s and $R^{(2)}_{3i}$'s are as defined in \eqref{rfnseom2} and \eqref{rfnseom2} respectively, and $L_{23}$ and $L_{35}$ are as defined in \eqref{l2}. Further, the terms with $\Theta(n-m-1)$ contribute only when $n\geq m+1$ and those with $\Theta(n-m-2)$ contribute only when $n\geq m+2$. The $f^{(1,p)}$'s are as obtained in \eqref{sol1} and \eqref{soln}, and they vanish if $p$ is not even
and non-negative.

The boundary condition \eqref{bc22} determines the solutions of eq. \eqref{e14} uniquely.
The solutions of $f^{(2, 2n, 2m)}_2$'s are
\begin{eqnarray}\label{sol22}
f^{(2,0,0)}_2 (\rho) &=& \frac{\rho^4}{6b^4} + \frac{4b^4 +\rho^4}{12b^4} \log\left(\frac{4b^4 - \rho^4}{4b^4 + \rho^4}\right) + \frac{4b^4 +\rho^4}{6b^4}\left(\log\left(\frac{4b^4 - \rho^4}{4b^4 + \rho^4}\right)\right)^2, \nonumber\\
f^{(2,2n,2m)}_2 (\rho) &=& \displaystyle\sum\limits_{a=0}^{3} t_{a} \int_0^\rho d\rho^{'}
\frac{W_{a}(\rho^{'})}{W(t_1, t_2, t_3)(\rho^{'})}, \ \text{for}  \ n\geq1.
\end{eqnarray}
Here $W(t_1, t_2, t_3)$ denotes the Wronskian of $t_a$'s defined in \eqref{t}. We recall
that the Wronskian is the determinant of the $3\times 3$ matrix whose $i$-th column
is the transpose of $(t_{i}, dt_{i}/d\rho, d^2 t_{i}/d\rho^2)$, for $i=1,2,3$. Here $W_a$ denotes the determinant of the same matrix with the $a$-th column replaced by the
transpose of $(0, 0, S^{(2,2n,2m)}_2 )$, where $S^{(2,2n,2m)}$ is as defined in \eqref{sc}.

The solutions  of $f^{(2,2n,2m)}_3$'s are
\begin{eqnarray}\label{sol23}
f^{(2,0,0)}_3 (\rho) &=& -\frac{(4b^4 -\rho^4)^2}{6b^4(4b^4+\rho^4)} + \frac{2}{3}\frac{(4b^4 -\rho^4)(4b^4 + 3\rho^4)}{(4b^4+\rho^4)^2} +\frac{(4b^4 -\rho^4)^2}{4b^4(4b^4+\rho^4)}\log\left(\frac{4b^4 - \rho^4}{4b^4 + \rho^4}\right) , \nonumber\\
f^{(2,2n,2m)}_3 (\rho) &=& \int_0^\rho d\rho^{'}
\ \frac{S^{(2,2n,2m)}_3 (\rho^{'})}{u(\rho^{'})}, \ \text{for}  \ n\geq1.
\end{eqnarray}
Here $u$ is as defined in \eqref{u} and $S^{(2,2n,2m)}$ is as defined in \eqref{sc}.

To summarize, the solutions \eqref{sol21}, \eqref{sol22} and \eqref{sol23} exactly capture the entire correction to the metric at second order, by specifying the coefficients of the Taylor series \eqref{f2i} of $f^{(2)}_i (\rho, \omega, \omega_1)$ in $\omega$ and $\omega_1$ about $\omega=\omega_1= 0$, recursively.

We can clearly extend our method to higher orders in the amplitude expansion, where we can solve the time dependence through similar recursion relations. 

\subsection{Subtleties in the Fourier space}
We have seen that order by order in the amplitude expansion, all the time derivatives
can be summed up efficiently in terms of a finitely few functions once we Fourier transform
the time dependence. The equations of motion, which are ODEs in $\rho$ with source terms, can also be obtained for these finitely few functions from the amplitude expansion of the tensor eq \eqref{ed} and Fourier transforming its time-dependence as well.

We will point out here that the Fourier transform involves some subtleties. These subtleties,
are however, not new and could be readily illustrated with Navier-Stokes equation with no
forcing term as an example. It is believed, though there is no rigorous proof, that the solutions will decay in the future and become homogeneous. Similarly, we believe in the case of AdS gravity, that the perturbations will decay and the solution will settle down to
a boosted black brane. The point is that such decaying functions, like $e^{-\alpha t}$ for instance, for $\alpha>0$, typically grow in the past and have no well-defined Fourier
transform by the standard contour prescription, where the integration in $t$ runs along the
real axis from $-\infty$ to $\infty$. 

The remedy for this is well known. It is convenient to state this remedy in terms of the
inverse Fourier transform. Let
\begin{equation}
f(\omega) = \frac{1}{2\pi }\left(\frac{1}{\alpha - i\omega}\right), \quad \alpha>0.
\end{equation} 
We see that $f(\omega)$ has a simple pole in the lower half plane at $\omega = -i\alpha$ with residue $-1/(2\pi i)$. Let us inverse Fourier transform $f(\omega)$ by doing this
integration over $\omega$ using the following contour. We go slightly below the real axis and integrate in $\omega$ from $-\infty-i\epsilon$ to $\infty+i\epsilon$, where $\epsilon$ is infinitesimal and positive, and then close the contour in the lower half plane by going along the semicircle at infinity. With this contour prescription, the inverse
Fourier transform picks up contribution only from the pole in the lower half plane, so
we get
\begin{equation}
f(t) = e^{-\alpha t},
\end{equation} 
which decays in the future.

This contour prescription is simple, but has an important physical significance. We have a similar well-known contour prescription when one defines the retarded correlator in real time. The physical significance is that we pick up an arrow of time. In the case of solutions of gravity
at the non-linear level, the result is even more drastic.

We know that for purely hydrodynamic solutions of gravity, we can either have a regular
future horizon or a regular past horizon, but not both \cite{myself1}. The former happens at the first order in the derivative expansion when $4\pi\eta /s= 1$, and the latter happens when $4\pi\eta/s=-1$. The regular perturbations of the AdS black brane and the AdS white brane cannot be glued without encountering some discontinuity, which is disallowed by Einstein's equation in vacuum with a negative cosmological constant. Therefore, the past of 
the solution which is purely hydrodynamic in the future will have drastically different 
behavior.

The same situation will be repeated here in the case of homogeneous relaxation.
In this case also, we will be interested in the solution which will settle to an AdS black brane in the future. Quite generally, the past behavior of the solution will depend on the 
specific initial condition. However, the behavior in the future can be expected to be governed by a generic phenomenological equation like the Navier-Stokes equation. provided the future horizon is regular. Remarkably, this generic behavior at late time is exactly what is expected also in the dual theory. Our contour prescription isolates the \textit{generic} future behavior from the \textit{non-generic} behavior in the past, the latter being dependent on the specific details of the initial state.

\section{Translation to ingoing Eddington-Finkelstein coordinate
system in the amplitude expansion}

The ingoing Eddington-Finkelstein coordinates are particularly
suitable for analyzing the future horizon. The position 
of the future horizon at late time is revealed in these 
coordinates manifestly. In the case of stationary black holes, the metric is also
manifestly regular at the future horizon in these coordinates.

For more general geometries, which are perturbations of AdS black branes, 
we need to give a definition of the ingoing Eddington-Finkelstein coordinate
system. This can be achieved by noting that the unperturbed boosted AdS black brane
takes the following covariant form 
\begin{equation}\label{EFBB}
ds^{2} = -2u_{\mu}dx^{\mu}dr + G_{\mu\nu}(r)dx^{\mu}dx^{\nu},
\end {equation}
where $u_{\mu}$ is the boost parameter of the black brane as before, and
\begin{equation}\label{covG}
G_{\mu\nu} = -r^2 \left(1- \frac{1}{b^4 r^4}\right)u_\mu u_\nu + r^2 P_{\mu\nu},
\end{equation}
with $P_{\mu\nu}$ being as in \eqref{P}, 
the projection tensor on the spatial hyperplane orthogonal to $u^\mu$. 

We can think of $r$ as the Eddington-Finkelstein radial coordinate and $x^{\mu}$'s as the boundary coordinates. The horizon can be located by finding where $G_{\mu\nu}u^\mu u^\nu$ has a simple zero. We can readily see that the horizon is located at $r=1/b$. Further the Hawking temperature $T = 1/(\pi b)$, so we can legally use the same notation $b$ as in the previous section, where $b$ was defined as $1/(\pi T)$. The generators of the horizon in these coordinates are also limits of radial outgoing null rays. So this horizon is a future horizon. We also note that if we have chosen the outgoing Eddington-Finkelstein coordinates, the $(\mu r)$ and $(r \mu)$ components of the metric would have been $u_\mu$ 
instead of $-u_\mu$, and the horizon would have been a past horizon.

The natural generalization of this form of the metric, to geometries which are
perturbations of the boosted black brane, is
\begin{equation}\label{EFG}
ds^{2} = -2u_{\mu}(x)dx^{\mu}dr + G_{\mu\nu}(r,x)dx^{\mu}dx^{\nu},
\end {equation}
where $u_\mu$ is now the $local$ boost parameter and $G_{\mu\nu}$ also depends
on the boundary coordinates $x^{\mu}$. As in \cite{myself1}, we define the ingoing Eddington-Finkelstein coordinate system as the system in which the metric assumes the above form. 

When $u^{\mu}$ is a constant four-vector, as in the previous section, it is convenient to go to the global comoving frame where the boost parameter $u^\mu$ is $(1,0,0,0)$. We define the 
ingoing Eddington-Finkelstein time $v$, as $v= -u_\mu x^\mu$ and call the spatial coordinates in the orthogonal spatial plane $x^i$'s, where $i=1,2,3$. The boosted black brane 
metric given by \eqref{EFBB} and \eqref{covG} now assumes the form
\begin{equation}\label{EFBBC1}
ds^{2} = 2dvdr + G^{(0)}_{00}(r)dv^2 + G^{(0)}_{ij}(r)dx^i dx^j,
\end {equation}
where,
\begin{equation}\label{comovG}
G^{(0)}_{00} = -r^2 \left(1- \frac{1}{b^4 r^4}\right), \quad G^{(0)}_{ij} = r^2 \delta_{ij}.
\end{equation}

In our case of homogeneous relaxation, the metric in the global comoving frame in this coordinate system should assume the form,
\begin{equation}\label{EFH}
ds^2 = 2dvdr + G_{00}(r,v) dv^2 + G_{ij}(r,v) dx^i dx^j, 
\end{equation}
where $G_{00}$ and $G_{ij}$ are also dependent on the time $v$. We note that $G_{i0} =G_{0i}= 0$ as in the case of $g_{i0}$ and $g_{0i}$ in the Fefferman-Graham
coordinate system, because there is no vector which can be constructed just from the 
boundary data $\pi_{ij}(t)$. We will soon show that this is indeed the case when we obtain
the metric explicitly by translating the Fefferman-Graham metric into this coordinate system.

\subsection{The equations for translation}

We have seen in the previous section that when the boundary stress-tensor assumes the form of homogeneous relaxation, the metric in Fefferman-Graham coordinates,
in the global comoving frame, summing over all orders in the amplitude expansion, assumes the form
\begin{equation}\label{FGH}
ds^2 = \frac{d\rho^2 + g_{00}(\rho, t)dt^2 + g_{ij}(\rho, t)dz^i dz^j }{\rho^2}
\end{equation}

If we do the following change of coordinates,
\begin{eqnarray}\label{tr1}
\rho &=& \Phi(r,v), \\\nonumber
t &=& v + k(r,v),\\\nonumber
z^{i} &=& x^{i},
\end{eqnarray}
the Fefferman-Graham metric \eqref{FGH} indeed translates to the ingoing Eddington-Finkelstein metric \eqref{EFH} provided
\begin{eqnarray}\label{tr2}
\left(\frac{\partial\Phi}{\partial r}\right)^{2} + g_{00}\left(\rho(r,v), t(r,v)\right)\left(\frac{\partial k}{\partial r}\right)^{2} &=& 0,\nonumber\\
\frac{\partial\Phi}{\partial r}\frac{\partial\Phi}{\partial v} + g_{00}\left(\rho(r,v), t(r,v)\right)\frac{\partial k}{\partial r}\left(1+\frac{\partial k}{\partial v}\right) &=& \Phi^2,
\end{eqnarray} 
where $g_{00}$ is now evaluated as a function of $r$ and $v$. These equations thus determine the change of coordinates \eqref{tr1} from the Fefferman-Graham coordinates
to the Eddington-Finkelstein coordinates.

Once the Fefferman-Graham coordinates are determined as functions of the ingoing Eddington-Finkelstein coordinates, one can readily translate the Fefferman-Graham metric to  the ingoing Eddington-Finkelstein metric. We note that, just like the Fefferman-Graham
metric \eqref{FGH} is fully specified by $g_{00}$ and $g_{ij}$, the ingoing Eddington-Finkelstein metric \eqref{EFH} is also fully specified by $G_{00}$ and $G_{ij}$. The change of coordinates \eqref{tr1}, determined by \eqref{tr2}, implies
\begin{eqnarray}\label{tr3}
\left(\frac{\partial\Phi}{\partial v}\right)^{2} + g_{00}\left(\rho(r,v),t(r,v)\right)\left(1+\frac{\partial k}{\partial v}\right)^2 &=& G_{00}\Phi^2 \nonumber\\
g_{ij}\left(\rho(r,v),t(r,v)\right) &=& G_{ij}\Phi^2.
\end{eqnarray}
The above equations thus translate the metric. We note that the equations above are algebraic equations and not PDEs like \eqref{tr2}, where $r$ and $v$ derivatives of the unknowns $\Phi$ and $k$ appear. In the above equations, the $r$ and $v$ derivatives of
the unknowns, $G_{00}$ and $G_{ij}$ do not appear.

For the purely hydrodynamic case too, similar equations like \eqref{tr2}  and \eqref{tr3} for changing coordinates and translating the metric respectively from the Fefferman-Graham to ingoing Eddington-Finkelstein system were derived \cite{myself1}. However, there the equations for changing coordinates and that for translating the metric did not disentangle nicely like here.  

For the unpertubed AdS black brane, $\Phi^{(0)}(r,v)$ and $k^{(0)}(r,v)$, which completely specify the change of coodinates \eqref{tr1} translating
the $g_{\mu\nu}$ in the Fefferman-Graham metric \eqref{zof} to 
the $G_{\mu\nu}$ in Eddington-Finkelstein coordinates \eqref{comovG} are, 
\begin{eqnarray}\label{zot}
\Phi^{(0)}(r,v) &=& \Phi^{(0)}(r) =
\frac{\sqrt{2}b}{\sqrt{b^{2}r^{2}+\sqrt{b^{4}r^{4}-1}}},\nonumber\\
k^{(0)}(r, v) &=& k^{(0)}(r) = b\left(\pi - 2 \ \arctan (br) +\frac{1}{4}
\log\left(\frac{br+1}{br-1}\right)\right).
\end{eqnarray}
We can readily see that these solve the eqs. \eqref{tr2}. Further, the eqs. \eqref{tr3} are also satisfied when $G^{(0)}_{00}$ and $G^{(0)}_{ij}$ are given by \eqref{comovG}. We also observe that
both $\Phi^{(0)}(r)$ and $k^{(0)}(r)$ behave like $1/r$, as we go towards the boundary at
$r=\infty$.

We will now see how to implement eqs. \eqref{tr2} and \eqref{tr3} for changing the coordinates and translating the metric respectively, from the Fefferman-Graham system to the ingoing Eddington-Finkelstein system, order by order in the amplitude expansion. At each order, we will
sum over all derivatives with respect to the Eddington-Finkelstein time $v$ through recursion
relations, just like we summed over all derivatives in Fefferman-Graham time $t$ in the previous section. 

\subsection{Translation at the first order in the amplitude expansion}

At each order in the amplitude expansion, it will be easier to solve the eqs. \eqref{tr2} specifying change of coordinates \eqref{tr1} first.  At the first order in the amplitude expansion, the change of coordinates do not receive any correction. We note that the change of coordinates from the Fefferman-Graham system to the ingoing Eddington-Finkelstein system, given by \eqref{tr1} maintains spatial translational and rotational symmetry of the former manifestly, to all orders in the amplitude expansion. It is easy to see that since $\pi_{ij}\delta_{ij}=0$, there is no scope to write anything which is invariant under spatial rotations and translations, that can possibly correct $\Phi(r,v)$ and $k(r,v)$, from their forms at the zeroth order given by \eqref{zot}. 

We recall that $g_{\mu\nu}$ in the Fefferman-Graham coordinates at the first order in the derivative expansion takes the form \eqref{fofs}, where $g^{(1)}_{00}$ vanishes and $g^{(1)}_{ij}$ is traceless. The equations for translation of the metric \eqref{tr3} at the first order in the amplitude
expansion therefore implies
\begin{eqnarray}\label{tr11}
G^{(1)}_{00} &=& 0, \nonumber\\
G^{(1)}_{ij} &=& \frac{g^{(1)}_{ij}\left(\rho = \Phi^{(0)}(r), t = v + k^{(0)}(r)\right)}{\left(\Phi^{(0)}(r)\right)^2},
\end{eqnarray}
where $\Phi^{(0)}(r)$ and $k^{(0)}(r)$ are as in \eqref{zot} for the unperturbed AdS black brane.

Further, just like $g^{(1)}_{ij}$ in \eqref{fofs},  we can write $G^{(1)}_{ij}$ as follows
\begin{equation}\label{tr12}
G^{(1)}_{ij} = \displaystyle\sum\limits_{n=0}^{\infty}b^{2+n}F^{(1,n)}(r)\left(\frac{d}{dv}\right)^n \pi_{ij}(v).
\end{equation}
The difference from \eqref{fofs} here is that $\pi_{ij}$ is a function of the Eddington-Finkelstein time $v$ and not a function of the Fefferman-Graham time $t$.  Additionally,  the time derivatives run over 
both even and odd powers, unlike even orders only in the latter case. Also, by definition,
$F^{(1,n)}$'s are dimensionless functions like $f^{(1,2n)}$s.

Substituting \eqref{fofs} and \eqref{tr12} in \eqref{tr11}, we get
\begin{eqnarray}\label{solefn}
F^{(1,n)}(r) &=& \frac{b^2}{\left(\Phi^{(0)}(r)\right)^2}\displaystyle\sum\limits_{m=0}^{\frac{n}{2}}f^{(1,2m)}\left(\rho = \Phi^{(0)}(r)\right)\frac{\left(k^{(0)}(r)\right)^{n-2m}}{b^{n-2m}(n-2m)!}\nonumber\\
&& \text{when}\ n \ \text{is even}, \nonumber\\
F^{(1,n)}(r) &=& \frac{b^2}{\left(\Phi^{(0)}(r)\right)^2}\displaystyle\sum\limits_{m=0}^{\frac{n-1}{2}}f^{(1,2m)}\left(\rho = \Phi^{(0)}(r)\right)\frac{\left(k^{(0)}(r)\right)^{n-2m}}{b^{n-2m}(n-2m)!}\nonumber\\
&& \text{when}\ n \ \text{is even}.
\end{eqnarray}
We recall that we have solved $f^{(1,2m)}$ in the previous section recursively and our solutions are given by \eqref{sol1} and \eqref{soln}. Thus the equations above also give us recursive solutions for $F^{(1,n)}(r)$. We will not make these recursion relations more explicit here. The leading term $F^{(1,0)}(r)$ in this recursion can be easily found to take the form
\begin{equation}\label{solef0}
F^{(1,0)}(r) = -b^2 r^2 \ \log\left(1-\frac{1}{b^4 r^4}\right).
\end{equation}
The recursion series \eqref{solefn} allows us to sum over all the time derivatives at the
first order in the amplitude expansion.

When we derived the recursion series in the Fefferman-Graham coordinate
system to sum over all time derivatives, we went to Fourier space. This has not been
necessary at the first order in the amplitude expansion here, because we had to only solve the
algebraic eq. \eqref{tr11}. At the second order in the amplitude expansion, the equations for
changing coordinates will be PDEs and not algebraic equations, so Fourier transforming the
time dependence will be necessary again.

Therefore, for the sake of completeness, let us see if the metric here also assumes a compact form here in Fourier space too. The Fourier transform of $\pi_{ij}(v)$ can be defined through
\begin{equation}\label{Fr2}
\pi_{ij}(v) = \int_{-\infty}^{\infty} d\omega \ e^{-i\omega v}\pi_{ij}(\omega).
\end{equation} 
We note that since $\pi_{ij}(v)$ is the same function of $v$ as $\pi_{ij}(t)$ is a function of $t$. Given that we defined the Fourier transform of $\pi_{ij}(t)$ in \eqref{Fr1} in the same way, it follows that the Fourier transforms $\pi_{ij}(\omega)$ are identical.

Similarly, we can define the Fourier transform of $G^{(1)}_{ij}(r, v)$ in $v$. Let us also define
$F^{(1)}(r, \omega)$ in analogy with $f^{(1)}(\rho,\omega)$ in \eqref{f1} ,as 
\begin{equation}\label{F1}
F^{(1)}(r,\omega) = \displaystyle\sum\limits_{n=0}^{\infty}(-i\omega b)^n F^{(1,n)}(r).
\end{equation}
Then we can indeed write $G^{(1)}_{ij}$ which gives the entire first order correction to the metric compactly in Fourier space as
\begin{equation}
G^{(1)}_{ij}(r,\omega) = b^{2}F^{(1)}(r,\omega)\pi_{ij}(\omega).
\end{equation}
This equation is analogous to \eqref{fof} where the full correction to the metric at the first
order has been captured by a single function $f^{(1)}(r,\omega)$ after Fourier transforming the time dependence in Fefferman-Graham coordinates.

Finally, the equation for translation \eqref{tr11} when Fourier transformed in $v$, gives
the following algebraic relationship between $F^{(1)}$ and $f^{(1)}$,
\begin{equation}\label{fourtr1}
F^{(1)}(r,\omega) = \frac{b^2}{\left(\Phi^{(0)}(r)\right)^2}f^{(1)}\left(\rho = \Phi^{(0)}(r), \omega\right)e^{-i\omega k^{(0)}(r)}.
\end{equation}
If we were able to solve the ODE for $f^{(1)}(\rho,\omega)$ given by \eqref{eom1} exactly, we could have easily used the algebraic equation above to know $F^{(1)}(r, \omega)$ exactly. However, this has not been possible, so we have taken care of the dependence on $\omega$, or equivalently summed the time derivatives recursively.  We have already translated the same strategy here in \eqref{solefn}.

\subsection{Translation at the second order in the amplitude expansion}

As in the case of the first order in the amplitude expansion, it is instructive to begin with how the changes in coordinates \eqref{tr1} take their
form at the second order in the amplitude expansion. Taking into account manifest
invariance under spatial translations and rotations, clearly $\Phi^{(2)}(r,v)$ and $k^{(2)}(r,v)$
specifying the change of coordinates at the second order in the amplitude expansion, should take the form,
\begin{eqnarray}
\Phi^{(2)}(r,v) &=& \displaystyle\sum\limits_{n=0}^{\infty}\displaystyle\sum\limits_{\substack{m=0\\n+m \ is  \ even}}^{n}  
b^{9+n}\Phi^{(2,n,m)}(r)\displaystyle\sum\limits_{\substack{a,b=0 \\ a+b=n \\ |a-b|=m}}^{n}\left(\frac{d}{dv}\right)^{a}\pi_{pq}(v)\left(\frac{d}{dv}\right)^{b}\pi_{pq}(v),\nonumber\\
k^{(2)}(r,v) &=& \displaystyle\sum\limits_{n=0}^{\infty}\displaystyle\sum\limits_{\substack{m=0 \\ n+m \ is \ even}}^{n} 
b^{9+n}k^{(2,n,m)}(r)\displaystyle\sum\limits_{\substack{a,b=0 \\ a+b=n \\|a-b|=m}}^{n}\left(\frac{d}{dv}\right)^{a}\pi_{pq}(v)\left(\frac{d}{dv}\right)^{b}\pi_{pq}(v),
\end{eqnarray}
where, by definition, $\Phi^{(2,n,m)}(r)$ and $k^{(2,n,m)}(r)$ are dimensionless functions, and $n+m$ is even ensures that the sum over $m$ is over even integers if $n$ is even, and is over odd integers if $n$ is odd, so that both $(n+m)/2$ and $(n-m)/2$, i.e. $a$ and $b$, are integers.

In Fourier space, we can put these changes of coordinates, compactly in terms of
just two functions, $\Phi^{(2)}(r,\omega, \omega_1)$ and $k^{(2)}(r, \omega, \omega_1)$,
through the following relation,
\begin{eqnarray}
\Phi^{(2)}(r,v) &=& b^9 \int_{-\infty}^{\infty} d\omega \ e^{-i\omega v}\int_{-\infty}^{\infty}d\omega_1  \ \Phi^{(2)}(r, \omega, \omega_1)\pi_{pq}(\omega_1)\pi_{pq}(\omega-\omega_1), \nonumber\\
k^{(2)}(r,v) &=& b^9 \int_{-\infty}^{\infty} d\omega \ e^{-i\omega v}\int_{-\infty}^{\infty}d\omega_1 \ k^{(2)}(r, \omega, \omega_1)\pi_{pq}(\omega_1)\pi_{pq}(\omega-\omega_1),
\end{eqnarray}
where,
\begin{eqnarray}\label{seriesef1}
\Phi^{(2)}(r,\omega,\omega_1) &=& \displaystyle\sum\limits_{n=0}^{\infty}\displaystyle\sum\limits_{\substack{m=0 \\ n+m \ is \ even}}^{n} (-ib)^n (\omega_1^{\frac{n+m}{2}}(\omega-\omega_1)^{\frac{n-m}{2}}\nonumber\\&&\qquad\qquad\qquad\qquad\qquad+\omega_1^{\frac{n-m}{2}}(\omega-\omega_1)^{\frac{n+m}{2}})\Phi^{(2,n,m)}(r), \nonumber\\
k^{(2)}(r,\omega, \omega_1) &=& \displaystyle\sum\limits_{n=0}^{\infty}\displaystyle\sum\limits_{\substack{m=0 \\ n+m \ is \ even}}^{n} (-ib)^n (\omega_1^{\frac{n+m}{2}}(\omega-\omega_1)^{\frac{n-m}{2}}\nonumber\\&&\qquad\qquad\qquad\qquad\qquad+\omega_1^{\frac{n-m}{2}}(\omega-\omega_1)^{\frac{n+m}{2}})k^{(2,n,m)}(r).
\end{eqnarray}
We note that, by definition $\Phi^{(2)}(r, \omega, \omega_1)$ and $k^{(2)}(r,\omega, \omega_1)$ are dimensionless functions.

We can readily obtain  the equations for $\Phi^{(2)}(r, \omega, \omega_1)$ by expanding the equations for change of coordinates \eqref{tr2} up to second order in the amplitude expansion, and then Fourier transforming the time dependence. Thus we have,
\begin{eqnarray}\label{tr21}
\frac{d \Phi^{(0)}(r)}{d r}\frac{\partial \Phi^{(2)}}{\partial r}(r,\omega, \omega_1)&+& g_{00}^{(0)}\left(\rho=\Phi^{(0)}(r)\right) \frac{d k^{(0)}(r)}{d r}\frac{\partial k^{(2)}}{\partial r}(r,\omega, \omega_1)\nonumber\\
&&+\frac{1}{2}\left(\frac{d g_{00}^{(0)}}{d\rho}\right)\left(\rho=\Phi^{(0)}(r)\right)\left(\frac{d k^{(0)}(r)}{d r}\right)^2 \Phi^{(2)}(r,\omega,\omega_1) \nonumber\\&&= -\frac{1}{2b}f^{(2)}_3 \left(\rho =\Phi^{(0)}(r), \omega, \omega_1\right)e^{-i\omega k^{(0)}(r)} \left(\frac{d k^{(0)}}{d r}\right)^2,\nonumber\\
g_{00}^{(0)}\left(\rho=\Phi^{(0)}(r)\right)\frac{\partial k^{(2)}}{\partial r}(r,\omega,\omega_1)&-&(i\omega\frac{d \Phi^{(0)}(r)}{d r} \nonumber\\&&- \left(\frac{d g_{00}^{(0)}}{d\rho}\right)\left(\rho=\Phi^{(0)}(r)\right)\frac{d k^{(0)}(r)}{d r} 
\nonumber\\&&+2\Phi^{(0)}(r)) \ \Phi^{(2)}(r,\omega,\omega_1)\nonumber\\&&-i\omega g_{00}^{(0)}\left(\rho=\Phi^{(0)}(r)\right) \frac{d k^{(0)}(r)}{d r}k^{(2)}(r,\omega,\omega_1)\nonumber\\
&&=-\frac{1}{b}f^{(2)}_3 \left(\rho =\Phi^{(0)}(r), \omega, \omega_1\right)e^{-i\omega k^{(0)}(r)} \frac{d k^{(0)}}{d r}\ .
\end{eqnarray}
The solutions are uniquely specified by the boundary conditions,
\begin{eqnarray}
\Phi^{(2)}(r,\omega,\omega_1) &=& O\left(\frac{1}{b^9 r^9}\right), \ at \ r\rightarrow\infty,
\nonumber\\
k^{(2)}(r,\omega,\omega_1) &=& O\left(\frac{1}{b^9 r^9}\right), \ at \ r\rightarrow\infty.
\end{eqnarray}
These boundary conditions follows from simple dimensional analysis. The leading behavior of $\Phi^{(0)}(r)$ and $k^{(0)}(r)$ at the boundary $r=\infty$ is $1/r$, and is independent of $b$. The leading bahavior of $\Phi^{(2)}(r,v)$ and $k^{(2)}(r,v)$ also should be independent of $b$, so dimensional analysis shows that their leading behavior should be $1/r^9$. This begets the above boundary conditions.

The eqs. \eqref{tr21} which are ordinary differential equations in $r$ are unfortunately not solvable for arbitrary $\omega$ and $\omega_1$. Therefore, we can adopt the same strategy as in the previous section, to solve these exactly first when $\omega=\omega_1 =0$, so that we know $\Phi^{(2,0,0)}(r)$ and $k^{(2,0,0)}(r)$ exactly and then sum over the dependence over $\omega$ and $\omega_1$ in \eqref{seriesef1} by obtaining $\Phi^{(2,n,m)}(r)$ and $k^{(2,n,m)}(r)$ recursively. We will not repeat this here because it will turn out that these exact recursion relations will not be important for the regularity analysis.

We also note that the eqs. \eqref{tr21} are first order ordinary differential equations in $r$. Given
that the equations for translation of the metric which we will derive next are algebraic, we bypass solving second order differential equations with constraints, involved in directly obtaining the metric in Eddington-Finkelstein coordinates from Einstein's equation. 

The $G_{\mu\nu}$ in the ingoing Eddington-Finkelstein metric \eqref{EFH}, similarly, should take the following form at the second order in the amplitude expansion,
\begin{eqnarray}
G^{(2)}_{00}(r,v) &=& \displaystyle\sum\limits_{n=0}^{\infty}\displaystyle\sum\limits_{\substack{m=0 \\ n+m \ is \ even}}^{n} 
b^{6+n}F^{(2,n,m)}_{3}(r)\displaystyle\sum\limits_{\substack{a,b=0 \\ a+b=n \\ |a-b|=m}}^{n}\left(\frac{d}{dv}\right)^{a}\pi_{pq}(v)\left(\frac{d}{dv}\right)^{b}\pi_{pq}(v), \nonumber\\
G^{(2)}_{ij}(r,v) &=&  \displaystyle\sum\limits_{n=0}^{\infty}\displaystyle\sum\limits_{\substack{m=0 \\ n+m \ is \ even}}^{n} 
b^{6+n}F^{(2,n,m)}_{2}(r)\delta_{ij}\displaystyle\sum\limits_{\substack{a,b=0 \\ a+b=n \\ |a-b|=m}}^{n}\left(\frac{d}{dv}\right)^{a}\pi_{pq}(v)\left(\frac{d}{dv}\right)^{b}\pi_{pq}(v) \nonumber\\
&&+\displaystyle\sum\limits_{n=0}^{\infty}\displaystyle\sum\limits_{\substack{m=0 \\ n+m \ is \ even}}^{n} 
b^{6+n}F^{(2,n,m)}_{1}(r)\displaystyle\sum\limits_{\substack{a,b=0 \\ a+b=n \\ |a-b|=m}}^{n}\Bigg[\left(\frac{d}{dv}\right)^{a}\pi_{ik}(v)\left(\frac{d}{dv}\right)^{b}\pi_{kj}(v)\nonumber\\
&&\qquad\quad\quad\qquad\qquad\qquad\qquad\quad -\frac{1}{3}\delta_{ij}\left(\frac{d}{dv}\right)^{a}\pi_{pq}(v)\left(\frac{d}{dv}\right)^{b}\pi_{pq}(v)\Bigg].
\end{eqnarray}
By definition, $F^{(2,n,m)}_i$s are dimensionless functions. Once again the sum over $m$ runs over even intergers if $n$ is even, and over odd integers if $n$ is odd.

After Fourier transforming the time dependence, in analogy with $g^{(2)}_{\mu\nu}(\rho, \omega, \omega_1)$ in the Fefferman-Graham metric at the second order in the amplitude expansion given by \eqref{sof}, $G^{(2)}_{\mu\nu}(r,\omega,\omega_1)$ can be compactly written in terms of three dimensionless functions $F^{(2)}_i (r,\omega, \omega_1)$, for $i=1,2,3$ as
\begin{eqnarray}
G^{(2)}_{00}(r,v) &=& b^6 \int_{-\infty}^{\infty} d\omega \ e^{-i\omega v}\int_{-\infty}^{\infty}d\omega_1 \ F^{(2)}_3(r, \omega, \omega_1)\pi_{pq}(\omega_1)\pi_{pq}(\omega-\omega_1), \nonumber\\
G^{(2)}_{ij}(r,v) &=& b^6  \int_{-\infty}^{\infty} d\omega \ e^{-i\omega v}\int_{-\infty}^{\infty}d\omega_1 \ F^{(2)}_2(r, \omega, \omega_1)\pi_{pq}(\omega_1)\pi_{pq}(\omega-\omega_1)\nonumber\\
&&+\frac{b^6}{2}  \int_{-\infty}^{\infty} d\omega \ e^{-i\omega v}\int_{-\infty}^{\infty}d\omega_1 \ F^{(2)}_1(r, \omega, \omega_1)[\pi_{ik}(\omega_1)\pi_{kj}(\omega-\omega_1)\nonumber\\
&&\qquad\quad\quad\qquad\qquad\qquad\qquad\qquad\qquad+\pi_{ik}(\omega-\omega_1)\pi_{kj}(\omega_1)\nonumber\\
&&\qquad\quad\quad\qquad\qquad\qquad\qquad\qquad\qquad-\frac{2}{3}\delta_{ij}\pi_{pq}(\omega_1)\pi_{pq}(\omega-\omega_1)],
\end{eqnarray}
where
\begin{eqnarray}\label{seriesef2}
F^{(2)}_i (r,\omega, \omega_1) &=& \displaystyle\sum\limits_{n=0}^{\infty}\displaystyle\sum\limits_{\substack{m=0 \\ n+m \ is \ even}}^{n} (-ib)^n (\omega_1^{\frac{n+m}{2}}(\omega-\omega_1)^{\frac{n-m}{2}}\nonumber\\&&\qquad\qquad\qquad+\omega_1^{\frac{n-m}{2}}(\omega-\omega_1)^{\frac{n+m}{2}})F^{(2,n,m)}_i (r), \text{for} \ i=1,2,3.
\end{eqnarray}

The equations for the translation of the metric \eqref{tr3} at the second order in the amplitude expansion, after Fourier transforming the time dependence implies,
\begin{eqnarray}\label{tr22}
F^{(2)}_3 (r,\omega, \omega_1) &=& \frac{b^2}{\left(\Phi^{(0)}(r)\right)^2}(f^{(2)}_3 \left(\rho=\Phi^{(0)}(r),\omega,\omega_1\right)e^{-i\omega k^{(0)}(r)}\nonumber\\&&\qquad\qquad \ \ +b\left(\frac{d g_{00}^{(0)}}{d\rho}\right)\left(\rho=\Phi^{(0)}(r)\right)\Phi^{(2)}(r,\omega,\omega_1)
\nonumber\\&&\qquad\qquad \ \ -i2b\omega g^{(0)}_{00} k^{(2)}(r,\omega, \omega_1)\nonumber\\&&\qquad\qquad \ \ -2b G_{00}^{(0)}(r)\Phi^{(0)}(r)\Phi^{(2)}(r,\omega,\omega_1)), \nonumber\\
F^{(2)}_2 (r,\omega, \omega_1) &=& \frac{b^2}{\left(\Phi^{(0)}(r)\right)^2}(f^{(2)}_2 \left(\rho=\Phi^{(0)}(r),\omega,\omega_1\right)e^{-i\omega k^{(0)}(r)}\nonumber\\&&\qquad\qquad \ \ +\frac{\left(\Phi^{(0)}(r)\right)^3}{b^3}\Phi^{(2)}(r,\omega,\omega_1)
\nonumber\\&&\qquad\qquad \ \ -2b r^2\Phi^{(0)}(r)\Phi^{(2)}(r,\omega,\omega_1)),\nonumber\\
F^{(2)}_1 (r,\omega, \omega_1) &=& \frac{b^2}{\left(\Phi^{(0)}(r)\right)^2} \ f^{(2)}_1 \left(\rho=\Phi^{(0)}(r),\omega,\omega_1\right)e^{-i\omega k^{(0)}(r)}.
\end{eqnarray}
Here we have used the explicit second order form of $g^{(2)}_{\mu\nu}(\rho,\omega, \omega_1)$ as given by \eqref{sof}. The above equations are algebraic. We recall that \eqref{eom21}, \eqref{eom22} and \eqref{eom23} gives the equations of motion for 
$f^{(2)}_i(\rho,\omega,\omega_1)$, for $i=1,2$ and $3$ respectively, and they have unique solutions given the boundary conditions \eqref{bc21} and \eqref{bc22}. If we use these and the solution for $\Phi^{(2)}(r,\omega,\omega_1)$, we obtain the full translation of the metric into the ingoing Eddington-Finkelstein coordinates through these algebraic equations.

In practice, we need to solve for the time dependence, or equivalently the $\omega$ and $\omega_1$ dependence of $F^{(2)}_i(r,\omega, \omega_1)$ for $i=1,2,3$ given by \eqref{seriesef2}, by solving the eqs. \eqref{tr22} for $\omega=\omega_1=0$ first, and then obtaining $F^{(2,n,m)}(r)$ and $k^{(2,n,m)}(r)$ recursively. We need to use the recursion relations for $f^{(2, 2n, 2m)}_i(\rho)$ given by
\eqref{sol21}, \eqref{sol22} and \eqref{sol23} for $i=1,2$ and $3$ respectively and also the recursion relations for $\Phi^{(2,n,m)}(r)$.

It will turn out that for the regularity analysis, we will only need recursion relations for $F^{(2,n,m)}_1(r)$ which, according to \eqref{tr2}, can be obtained without knowing $\Phi^{(2,n,m)}(r)$. We get the recursion relations,
\begin{eqnarray}\label{recef}
F^{(2,n,m)}_1 (r, \omega, \omega_1) &=& \frac{b^2}{2\left(\Phi^{(0)}(r)\right)^2}\displaystyle\sum\limits_{p=0}^{\frac{n}{2}}\Bigg(\displaystyle\sum\limits_{q=0}^{m}f^{(2,2p, m-q)}_1\left(\rho = \Phi^{(0)}(r)\right)\nonumber\\&&+\displaystyle\sum\limits_{q=0}^{2p -m}f^{(2,2p, m+q)}_1\left(\rho = \Phi^{(0)}(r)\right)\Bigg)\frac{\left(k^{(0)}(r)\right)^{n-2p}}{b^{n-2p}q!(n-2p -q)!}, \nonumber\\&&\ \text{when} \ n \ \text{is even}, \nonumber\\
F^{(2,n,m)}_1 (r, \omega, \omega_1) &=& \frac{b^2}{2\left(\Phi^{(0)}(r)\right)^2}\displaystyle\sum\limits_{p=0}^{\frac{n-1}{2}}\Bigg(\displaystyle\sum\limits_{q=0}^{m}f^{(2,2p, m-q)}_1\left(\rho = \Phi^{(0)}(r)\right)\nonumber\\&&+\displaystyle\sum\limits_{q=0}^{2p -m}f^{(2,2p, m+q)}_1\left(\rho = \Phi^{(0)}(r)\right)\Bigg)\frac{\left(k^{(0)}(r)\right)^{n-2p}}{b^{n-2p}q!(n-2p -q)!}, \nonumber\\&&\ \text{when} \ n \ \text{is odd},
\end{eqnarray}
where in turn $f^{(2,r,s)}_1(\rho)$ are defined through the recursion relations \eqref{sol21},
and vanish unless both $r$ and $s$ are even by their definitions in \eqref{f2i}. As a specific instance,
\begin{equation}
F^{(2,0,0)}_1 (r) = \frac{b^2 r^2}{2} \ \left(\log\left(1-\frac{1}{b^4 r^4}\right)\right)^2,
\end{equation}
which is the only contribution to $F^{(2)}_1$ if $\pi_{ij}$ is time-independent.

Clearly, we can use our method for translating the metric into ingoing Eddington-Finkelstein coordinates from the Fefferman-Graham coordinates at higher orders in the amplitude expansion, summing over all time derivatives at each order.

\section{Regularity analysis of the metric and new phenomenological parameters}

As discussed in the Introduction, it can be expected
that the metric will be manifestly regular at the future horizon
in an appropriate coordinate system order by order in the
amplitude expansion provided we sum over all time
derivatives. The amplitude of the non-hydrodynamic shear-stress
tensor becomes arbitrarily small when compared to the pressure
close to equilibrium. However, its time derivatives are not small
compared to the temperature even close to equilibrium. This
necessitates summing over all time derivatives while
treating the amplitude expansion perturbatively.

When the energy-momentum tensor is purely hydrodynamic, the
derivatives of the hydrodynamic variables are small compared to
the local temperature. The derivative expansion, therefore, can
be employed perturbatively. This logic was instrumental in the
construction of the metric perturbatively in the derivative
expansion, which was regular at the future horizon for
appropriate choice of the transport coefficients. We expect that
in our case, similarly, we will be able to extract an equation
of motion for the shear-stress tensor order by order in the
amplitude expansion, so that when the shear-stress tensor
follows this equation of motion with appropriate values of the
phenomenological parameters, the metric will be regular order by
order in the amplitude expansion. In this Section, we will
outline the details of this analysis.

Before we go through the details, we need to consider if we can do the regularity analysis more
efficiently by calculating curvature invariants like $R^{\mu\nu\rho\sigma}R_{\mu\nu\rho\sigma}$ at the horizon. Unlike the hydrodynamic case, where the regularity simply involves the transport coefficients, and regularity analysis simply tells us we have a regular horizon provided some algebraic equations involving the transport coefficients are satisfied, we expect that the regularity condition here will involve an equation of motion for $\pi_{ij}(t)$. Calculation of curvature invariants, which we will not reproduce here, also indicates so. However, a curvature invariant can at best reproduce the trace of the equation of motion for $\pi_{ij}(t)$, with for instance, a linear combination of $\pi_{ij}(t)$ and its time-derivatives. Therefore, one needs to compare several curvature invariants, to extract the equation of motion itself. This has proved to be a very difficult task when we have tried to do it. More than this practical issue, one cannot be sure of the regularity unless one has compared all possible curvature invariants and this requires utilising geometric properties of the spacetime. It seems that it is very difficult to do this also. The most elegant way to do the regularity analysis, it appears, is to go to a coordinate system where the metric can be argued to be manifestly regular. Here, we will show that we can indeed do so in the ingoing Eddington-Finkelstein coordinates.

We will first do the regularity analysis at the linear order,
that is at the first order in the amplitude expansion. Then we
will extend our analysis to higher orders in the amplitude
expansion. This will allow us to extract the equation of motion
for $\pi_{ij}(t)$ in the amplitude
expansion, such that when this is obeyed, the corresponding
metric in the bulk has a regular future horizon. Finally, we
will connect our results with our previous work and see how we
can fix some more general phenomenological coefficients at strong coupling.

\subsection{Regularity at the linear order}

At the level of linearized gravity perturbations with right
asymptotic behavior, quasinormal modes are those which are
ingoing at the future horizon. Though there is no
rigorous proof, there are strong arguments 
\cite{Horowitz, Berti} to suggest that
for linearized gravity perturbations, the ingoing boundary
condition at the future horizon is equivalent to manifest
regularity of the metric at the future horizon in ingoing
Eddington-Finkelstein coordinates. In other words, when the
metric is perturbed with the quasinormal modes, not only
combinations of the linear perturbations of the
metric invariant under linearized diffeomorphisms, but the full metric tensor itself will still be
manifestly regular at the future horizon in ingoing
Eddington-Finkelstein coordinates.

At the first order in amplitude expansion, the dependence on the
perturbation of the non-hydrodynamic shear-stress tensor is
linear, so the manifest regularity at the future horizon in
ingoing Eddington-Finkelstein coordinates should require that
the shear-stress tensor has time dependence given by a linear
superposition of the quasinormal mode frequencies at zero
wavelengths.

In the computations of the quasinormal modes for $AdS$ black
branes, usually the ingoing boundary condition is imposed first and
the dispersion relation is then extracted by requiring that the
Dirichlet boundary condition is obeyed at asymptotic spatial infinity. This
would imply that the non-normalizable mode of the perturbation
vanishes. However, in the approach employed here, we first find
the metric corresponding to an arbitrary $\pi_{ij}(t)$ in a flat
background in the dual theory, thus fixing both the
non-normalizable and the normalizable modes of the metric
perturbations. These determine the solution uniquely. The
quasinormal modes can be expected to be extracted by requiring
manifest regularity of the metric at the future horizon in
ingoing Eddington-Finkelstein coordinates. Our method can be readily generalized at the
non-linear level, besides it will give us the most general metrics with regular future horizons
dual to homogeneous relaxation.

To see how this works, it will be instructive to  
review the case of the purely hydrodynamic shear-stress tensor in this new light.

Let us consider a fluid at constant four-velocity $u^{(0)\mu}$
and constant temperature $T^{(0)}$. It is convenient to define
$b^{(0)}$ as $1/(\pi T^{(0)})$ as before, so that$1/b^{(0)}$ is the
radial location of the future horizon in the dual geometry in
ingoing Eddington-Finkelstein coordinates.
Further we define the projection tensor in the spatial plane
orthogonal to $u^{(0)\mu}$ as
\begin{equation}
P^{(0)\mu\nu}= u^{(0)\mu}u^{(0)\nu}+\eta^{\mu\nu}.
\end{equation}
The most general purely hydrodynamic shear-stress tensor linear
in the fluctuation in velocity, $\delta u^{\mu}$ and the
temperature, $\delta T$, is
\begin{eqnarray}\label{fluc}
t_{\mu\nu} &=&
\frac{4u^{(0)}_{\mu}u^{(0)}_{\nu}+\eta_{\mu\nu}}{4b^{(0)4}} -
\frac{4u^{(0)}_{\mu}u^{(0)}_{\nu}+\eta_{\mu\nu}}{b^{(0)4}}\left(\frac{\delta
b}{b^{(0)}}\right) +\frac{u^{(0)}_{\mu}\delta u_{\nu}+\delta
u_{\mu}u^{(0)}_{\nu}}{b^{(0)4}}\nonumber\\
&&-\frac{\gamma}{2b^{(0)3}}\sigma _{\mu\nu}+ O(\epsilon^2 ), \ \text{with} \\
\sigma_{\mu\nu}&=&\left(
\left(P^{(0)\alpha}_{\mu}P^{(0)\beta}_{\nu}\right)\frac{(\partial_{\alpha}\delta
u_{\beta} +\partial_{\beta}\delta u_{\alpha})}{2} -
\frac{1}{3}P^{(0)}_{\mu\nu}(\partial_\alpha \delta u^\alpha
)\right),
\end{eqnarray}
where $O(\epsilon^2 )$ denote terms at second and higher orders
in the derivative expansion. Also $\gamma$ is a dimensionless
quantity defined through
\begin{equation}
\gamma = \frac{4\pi\eta}{s},
\end{equation}
with $\eta$ being the shear viscosity and $s$ the entropy
density.

The corresponding metric in ingoing Eddington-Finkelstein
coordinates up to first order in the derivative expansion is,
\begin{eqnarray}
ds^2 &=& - 2 (u^{(0)}_{\mu} + \delta u_{\mu}) dx^\mu dr +
G_{\mu\nu} dx^{\mu}dx^{\nu}, \\\nonumber
G_{\mu\nu} &=& r^2 P^{(0)}_{\mu\nu} + (-r^2 +
\frac{1}{b^{(0)4}r^2}) u^{(0)}_{\mu}u^{(0)}_{\nu} -
\frac{4}{b^{(0)4}r^2}\left(\frac{\delta b}{b^{(0)}}\right)
u^{(0)}_{\mu}u^{(0)}_{\nu}\\\nonumber
&&+\frac{1}{b^{(0)4}r^2} \left(u^{(0)}_{\mu}\delta u_{\nu}+\delta
u_{\mu}u^{(0)}_{\nu}\right) \\\nonumber&& -
r\left(u^{(0)}_{\mu}(u^{(0)\alpha}\partial_\alpha )\delta u_\nu
+ u^{(0)}_{\nu}(u^{(0)\alpha}\partial_\alpha )\delta u_\mu
-\frac{2}{3}u^{(0)}_{\mu} u^{(0)}_{\nu}(\partial^\alpha \delta
u_\alpha )\right)\\\nonumber
&&+\left(2b^{(0)}r^2 H\left(b^{(0)}r\right) + \frac{(\gamma
-1)}{4} b^{(0)}r^2 \
\log\left(1-\frac{1}{b^{(0)4}r^4}\right)\right)\sigma_{\mu\nu}
+O\left(\epsilon^2 \right),
\end{eqnarray}
where $\sigma_{\mu\nu}$ is as defined in \eqref{fluc}, $O(\epsilon^2 )$
denotes terms at higher orders in the derivative expansion, and
\begin{equation}
H(x) = \frac{1}{4}\left( \log \left(\frac{(x+1)^2 (x^2
+1)}{x^4}\right)- 2 \ \arctan(x) +\pi \right).
\end{equation}
The metric above is just the linearization of the metric for arbitrary $\gamma$ in \cite{myself1}, 
which generalizes the metric in \cite{Bhattacharyya1} for $\gamma=1$.

We see that corrections to $G_{\mu\nu}$ in the metric in ingoing
Eddington-Finkelstein coordinates, which is linear in $\delta
u^\mu$ and $\delta T$, can be decomposed into four categories
based on whether the components are in the direction of the flow
$u^{(0)\mu}$ or orthogonal to it. The first category is
longitudinal-longitudinal, i.e. proportional to
$u^{(0)}_{\mu}u^{(0)}_{\nu}$; the second category is
longitudinal-transverse , i.e proportional to terms like
$u^{(0)}_{\mu} \delta u_\nu$, or $u^{(0)}_{\mu}(u^{(0)}\cdot
\partial) \delta u_\nu$; the third category is pure trace and
transverse-transverse, i.e. proportional to $P^{(0)}_{\mu\nu}$;
and the fourth one is traceless and transverse-transverse, i.e
proportional to terms like $\sigma_{\mu\nu}$. This division can
obviously be done at any order in the derivative expansion.

We also see that, at the first order in the derivative expansion, the
\textit{leading} divergence at the late-time horizon which
coincides with the unperturbed horizon $r= 1/b^{(0)}$ is
$((\gamma -1)/4) \log (b^{(0)}r -1)\sigma_{\mu\nu}$ and belongs
to the fourth category, which is traceless and transverse-transverse.
This divergence vanishes provided $\gamma = 1$, i.e. when
$\eta/s = 1/4\pi$. The other divergences are of the same
category, and are coefficients of $(b^{(0)}r-1)\log(b^{(0)}r-1)$ and
$(b^{(0)}r-1)^2 \log(b^{(0)}r-1)$, both of which are proportional
to $(\gamma-1)\sigma_{\mu\nu}$, which vanish for the same choice
of $\eta/s$. These terms are potentially divergent at the
horizon because their first or second derivatives, or both, with respect
to $r$ diverges at the horizon, i.e. at $r=1/b^{(0)}$.

If we do not fix $\gamma$ (or equivalently $\eta/s$) at the first order and obtain the metric at higher orders in the derivative expansion, we find various
other kinds of divergence in \textit{all} four categories of
$G_{\mu\nu}$ in the ingoing Eddington-Finkelstein metric. When $\gamma = 1$ all
divergences in \textit{all} four categories of $G_{\mu\nu}$,
with \textit{higher} powers of the logarithm, like
$\left(\log(b^{(0)}r-1)\right)^2$ and proportional to derivatives of $\sigma_{\mu\nu}$ vanish. However, only three divergences of lower orders remain.

These are as follows, one is a $\log(b^{(0)}r-1)$
divergence, another is a $(b^{(0)}r-1)\log(b^{(0)}r-1)$ divergence and the other one is a $(b^{(0)}r-1)^2\log(b^{(0)}r-1)$, all of which at the second order in the derivative expansion, for instance, are proportional to $(u^{(0)}\cdot\partial)\sigma_{\mu\nu}$ (which is Weyl covariant at linear order in the velocity fluctuation $\delta u$), and belongs to the traceless and transverse-transverse category as it is orthogonal to $u^\mu$. 
The cancellation of these divergences requires fixing the only
linear transport coefficient at the second order in the
derivative expansion in a flat background, whose contribution to
the shear-stress tensor at the second order in the derivative
expansion, is proportional to
$(u^{(0)}\cdot\partial)\sigma_{\mu\nu}$. Further, all these
divergences are proportional to a polynomial which is linear in the arbitrary choice of this second
order transport coefficient and vanish when this transport coefficient
takes the same value, which is $(2 - \log 2)/b^{(0)2}$.

It can be shown that the feature just mentioned persists up to
arbitrary higher orders in the derivative expansion. The
transport coefficients at any order can always be fixed by
taking care of the $\log(b^{(0)}r -1)$ divergence in the
traceless and transverse-transverse part of $G_{\mu\nu}$, and
once these are fixed at any given order, there is no other kind
of divergence in any category of $G_{\mu\nu}$ provided the
transport coefficients at \textit{all lower} orders have been
chosen correctly. Also this $\log(b^{(0)}r-1)$ divergence is
linear in the arbitrary choices of the transport coefficients at
that order. To summarize, the cancellation of the
$\log(b^{(0)}r-1)$ divergence ensures that all other divergences
vanish order by order in the derivative expansion, but not
vice-versa.

\textit{Therefore, the regularity condition at the future
horizon is simply that the coefficient of $\log(b^{(0)}r-1)$ term
in the traceless and transverse-transverse part of $G_{\mu\nu}$
should vanish, and this fixes the value of all transport
coefficients to all orders in the derivative expansion.} Up to
second order in the derivative expansion, explicit calculations
have shown that the values of transport coefficients determined
by regularity at the future horizon in ingoing
Eddington-Finkelstein coordinates match with those obtained from
the dispersion relations of the hydrodynamic shear and sound
quasinormal modes.

Even in the case of homogeneous relaxation, the regularity
condition at the horizon should be the same. The argument for it
is easy. The regularity condition should be a condition on
$\pi_{ij}(t)$, therefore it is independent of $r$. If the
divergence at the late-time horizon has to vanish at all times,
the coefficient of each singular term in the series expansion of
$G_{\mu\nu}$ in $r$ has to vanish at $r=1/b$, where the future
horizon is located at late time. \footnote{We recall that b is constant here to all
orders in the amplitude expansion.} The singular terms should
also appear in the traceless and transverse-transverse part of
$G_{\mu\nu}$, since all the corrections to the metric at the
first-order in the amplitude expansion being porportional to
$\pi_{ij}(t)$ and its time derivatives as shown in \eqref{tr11} and \eqref{tr12}, are traceless and
transverse-transverse.

Further, the singular term in the series expansion of
$G_{\mu\nu}$ in $r$ at $r=1/b$, whose coefficient should give a sufficient
condition for all other divergent terms in $r$ to vanish at the late time horizon, can be identified
as the $\log(br-1)$ term, by looking at the case when
$\pi_{ij}$ is time-independent. In this case, the entire contribution to the metric at the first order
in the amplitude expansion, according to \eqref{tr12} and \eqref{solef0} is given by
\begin{equation}
G^{(1)}_{ij}(r) = b^2 F^{(1,0)}(r) \pi_{ij} = -b^4 r^2 \ \log\left(1-\frac{1}{b^4 r^4}\right)\pi_{ij}.
\end{equation}
Indeed the vanishing of the $\log(br-1)$ term, which is the leading divergence at $r=1/b$ here, gives the sufficient condition $\pi_{ij}=0$, for regularity. This is expected because the black hole
cannot support any tensor hair.

Time-dependence of $\pi_{ij}(t)$
brings in divergences with arbitrary higher powers of the
logarithm, i.e. terms like $\left(\log(br-1)\right)^n$, where
$n>1$ is an integer; however it is the time-dependent
corrections to the coefficient of $\log(br-1)$ that should
control all the other divergences. In fact, simple inspection   
of $F^{(1)}(r,\omega)$ through \eqref{fourtr1}, 
shows that the coefficient of $\log(br-1)^n$ term for $n\geq2$ is $O(\omega^n)$.
As these coefficients of such singular
terms have higher powers of $\omega$ at the leading order itself, these should be total time derivatives of the coefficient of the
$\log(br-1)$ term, which is $O(1)$ in the $\omega$ expansion. Thus,
the vanishing of the latter can ensure that other divergences
vanish, but not vice-versa.

Therefore, \textit{the regularity condition at the future
horizon is that the coefficient of the $\log(br-1)$ term in the
traceless and transverse-transverse part of $G_{\mu\nu}$ in the
ingoing Eddington-Finkelstein metric should vanish, which
implies that $\pi_{ij}(t)$ should obey a linear differential
equation that can be calculated from Einstein's equation.}

If the perturbations corresponding to quasinormal modes
are indeed manifestly regular in the ingoing
Eddington-Finkelstein coordinates, by the argument above, the 
only possibility for the regularity condition at the
horizon is that the coefficient of the $\log(br-1)$ term should vanish, so that
it will also provide a sufficient condition for all other divergences to vanish.

We need to understand now how we can reproduce the quasinormal
mode frequencies at zero wavelengths from this regularity
condition. Let the coefficient of the $\log(br-1)$ term in 
the series expansion in $r$ of $F^{(1)}(r,\omega)$ about $r=1/b$ be $D_R^{(1)}(\omega)$. Then
our regularity condition at the linear order is
\begin{equation}\label{rc1}
C_{R \ ij}(v) =b^2 \int_{-\infty}^{\infty}d\omega \ e^{-i\omega
v}D_{R}^{(1)}(\omega)\pi_{ij}(\omega) = 0,
\end{equation}
where the integration is done with the contour prescription
mentioned before, which picks up contributions only from the
poles of the analytic continuation of
$D_{R}^{(1)}(\omega)\pi_{ij}(\omega)$ as a function of the
complex variable $\omega$ in the lower half plane. By definition, $D_R^{(1)}(\omega)$
is a dimensionless function. We have also observed before
that $\pi_{ij}(\omega)$ is identical if it is the Fourier transform of $\pi_{ij}(t)$ or
$\pi_{ij}(v)$, because the latter are the same functions, though of different variables. So,
the regularity condition on $\pi_{ij}(\omega)$ leads directly to the regularity condition on
$\pi_{ij}(t)$ after doing Fourier transform in $t$.

Further, near $\omega =0$, we can write,
\begin{equation}\label{ps}
D_{R}^{(1)}(\omega) = \displaystyle\sum\limits_{n=0}^{\infty}D_R^{(1,n)}
(-ib\omega)^n,
\end{equation}
where clearly $D_R^{(1,n)}$ is the coefficient of $\log(br-1)$ term in the series expansion of 
$F^{(1,n)}(r)$ about $r=1/b$ as is obvious from \eqref{F1}. By definition, $D^{(1,n)}$'s are dimensionless numbers which can be explicitly evaluated using \eqref{solefn}. For instance,
\begin{equation}
D_R^{(1,0)}= -1, \quad D_R^{(1,1)}=-(\pi/2) - (1/4) \ \log(2), \quad etc.
\end{equation}

However, this power series \eqref{ps} can have a radius of convergence. So, it is
necessary to define an analytic continuation of
$D_{R}^{(1)}(\omega)$ beyond this radius of convergence in the
lower half plane, for doing the Fourier transform using the mentioned contour prescription. It is
also important to realize that it is the \textit{analytic
continuation} of $D_{R}^{(1)}(\omega)$ and $\pi_{ij}(\omega)$ in
the lower half plane as a function of the complex variable
$\omega$ which is required to make connection with the
quasinormal modes. 

If the analytic continuation of $D_{R}^{(1)}(\omega)$ have
simple zeroes at $\pm \ \omega_{R_n}- i \omega_{I_n}$, with
$\omega_{I_n}>0$, in the lower half plane, and 
$\pi_{ij}(\omega)$ have
at most simple poles at $\pm \ \omega_{R_n}- i \omega_{I_n}$ in the lower half plane, then it follows that $D_{R}^{(1)}(\omega)\pi_{ij}(\omega)$ is analytic in the
lower-half plane and its Fourier transform vanishes by our
contour prescription as required by the regularity condition \eqref{rc1}. Thus
the regularity condition implies that in the lower half complex $\omega$ plane,
$\pi_{ij}(\omega)$ should have the following general form
\begin{equation}
\pi_{ij}(\omega) = \displaystyle\sum\limits_{n=0}^{\infty} \left(\frac{1}{2\pi}\right)\left(\frac{a_{n \
ij}}{\omega_{I_n}-i(\omega-\omega_{R_n})}+\frac{a_{n \
ij}^*}{\omega_{I_n}-i(\omega+\omega_{R_n})}\right) +
f(\omega),
\end{equation}
where $f(\omega)$ for large values of $|\omega|$ decays like
$O(1/|\omega|^2)$. So for $t>0$, the
Fourier transform of $\pi_{ij}(\omega)$ is
\begin{equation}
\pi_{ij}(t) = \displaystyle\sum\limits_{n=0}^{\infty} \left( a_{n \ ij} e^{-i
\omega_{R _n }t} +a_{n \ ij}^* e^{i \omega_{R _n}t} \right)
e^{- \omega_{I_n}t},
\end{equation}
which will match with the prediction of the quasinormal modes if
$\pm \ \omega_{R_n}- i \omega_{I_n}$, with $\omega_{I_n}>0$ are the
quasinormal mode frequencies in the scalar channel when the
wavelengths vanish. This spectrum in the scalar channel has been numerically
calculated in \cite{Starinets, Kovtun}. \footnote{In \cite{Starinets}, it was also found that the quasinomal spectrum at zero wavelengths in the scalar channel is well approximated by $2\pi Tn\left(\pm 1 - i\right)$ for large $n$.}

The coefficients of other singular terms, like $\log(br-1)^2$,
for instance, in $G_{\mu\nu}$ in ingoing Eddington-Finkelstein
metric at this order, should also take the form,
\begin{equation}
\int_{-\infty}^{\infty}d\omega \ e^{-i\omega
v}D^{(1)}(\omega)\pi_{ij}(\omega) = 0,
\end{equation}
where $D^{(1)}(\omega)$ is the coefficient of the $\log(br-1)^2$ term in the
of $F^{(1)}(r,\omega)$ obtained in the
previous section. We require that the analytic continuation of
$D^{(1)}(\omega)$ also should have
simple zeroes at the same points $\pm \ \omega_{R_n}- i
\omega_{I_n}$ in the lower half complex plane. We recall that, in
the purely hydrodynamic case, for instance, at the second order in
the derivative expansion, the analogous coefficient was a second
order polynomial in $\gamma$ (i.e. $4\pi\eta/s$), which had a simple zero at $\gamma =1$, just like the
coefficient of the $\log(br-1)$ term. Thus the regularity condition \eqref{rc1} should 
provide sufficient condition for cancellation of all divergent terms.

To summarize, we see that at the first order in the amplitude expansion,
regularity analysis at the late-time horizon simply implies that
the metric perturbations are composed of the infinite tower of
quasinormal modes in the scalar channel with zero wavelengths.

To check this explicitly, we need to find the
power series \eqref{ps} defining $D^{(1)}_R (\omega)$ about $\omega=0$ 
explicitly. We have been able to define this implicitly as $D_R^{(1,n)}$ are the coefficients of 
the $\log (br-1)$ terms of $F^{(1,n)}(r)$, which have been obtained through the recursion relation
\eqref{solefn}. If we can find an explicit recursion relation in terms of $D_R^{(1,n)}$
directly, we will be able to see how we can analytically continue the power series
\eqref{ps} in the lower half complex $\omega$ plane, and check explicitly if it has only simple zeroes and if they match with the quasinormal spectrum. This explicit recursion relation is unfortunately a hard combinatorial problem and we have been unable to solve it here.

\subsection{The full regularity analysis}

When the shear-stress tensor is purely hydrodynamic, the metric
continues to be manifestly regular at the late-time horizon in
the ingoing Eddington-Finkelstein coordinates at the non-linear
level too, order by order in the derivative expansion. It is
this feature that allows us to determine the non-linear
transport coefficients which do not affect the dispersion
relations of the
hydrodynamic shear mode and sound modes, order by order in the
derivative expansion. At second order in the derivative
expansion, in fact, we have three such non-linear transport
coefficients, whose values have been determined by this method \cite{Bhattacharyya1}.

An arbitrary quasinormal mode, even if it is non-hydrodynamic,
can be expected to be manifestly regular at the future horizon
in ingoing Eddington-Finkelstein coordinates. However, at the
non-linear level, when we build solutions corresponding to a
"liquid" of such fluctuations, we have no strong general
argument why the solution should be manifestly regular at the
late-time horizon in ingoing Eddington-Finkelstein coordinates.

Fortunately, in our present case of homogeneous relaxation we can
strongly argue that the regularity of the metric at the
late-time horizon will still be manifest in the ingoing
Eddington-Finkelstein coordinates, order by order in the
amplitude expansion.

The metric in the Fefferman-Graham
coordinates always maintains the symmetries of the dual boundary
configuration, in this case the symmetry under spatial
translations and rotations. However, this metric is not
manifestly regular at the late-time horizon even when the
regularity condition is satisfied. The change of coordinates
which will achieve the manifest regularity at the late-time
horizon can also be expected to respect the symmetries of the
dual boundary configuration. This means that the spatial
Fefferman-Graham coordinates $z^i$'s should translate to the new
spatial coordinates $x^i$'s in order to maintain manifest
spatial translation and rotational invariance up to rotations
and shifts which depend on the new radial coordinate $r$ and new
time coordinate $v$ only. So,
\begin{equation}
z^{i} = R^{i}_{\phantom{i}j}(r,v)x^j + S^{i}(r,v).
\end{equation}
However, it is not possible to write an orthogonal matrix
$R^i_{\phantom{i}j}$ and a shift vector $S^{i}$ using just
$\pi_{ij}$ and its derivatives. The rotational matrix and the
shift vector, therefore are constants, in which case of course
the metric remains invariant.

Further, to maintain manifest spatial translational and
rotational symmetries, we obviously require that the
Fefferman-Graham radial coordinate $\rho$ and time-coordinate
$t$, should be translated to new radial coordinate $r$ and time
coordinate $v$ without involving the spatial coodinates, so that
both $\rho$ and $t$ will be functions of $r$ and $v$ only.
Besides, we want manifest regularity only at the future horizon
and not at the past horizon. So the obvious choice is the
ingoing Eddington-Finkelstein gauge where $g_{rv}=g_{vr}=1$
(rather than -1) and $g_{rr}= g_{ri} = 0$, as at the zeroth
order this choice makes the regularity at the future horizon
manifest. We can expect this choice to achieve manifest
regularity at the future horizon to all orders in the amplitude
expansion.

Repeating the analysis at the linear level or at the first order
in the amplitude expansion, the regularity should be guaranteed
by the vanishing of a certain singular term in the series
expansion in $r$ of $G_{\mu\nu}$ in the ingoing
Eddington-Finkelstein coordinates, otherwise we cannot guarantee
regularity at all times. Further, the choice of this singular
term can be made by doing the regularity analysis at the linear
level, so \textit{the regularity condition up to any given order
in the amplitude expansion is the vanishing of the coefficient
of the $\log(br-1)$ term in the traceless and
transverse-transverse part of $G_{\mu\nu}$ in the ingoing
Eddington-Finkelstein metric}.

Clearly, the regularity condition up to second order in the
amplitude expansion can be written in the form,
\begin{eqnarray}\label{ft2}
C_{R \ ij}(t) &=& b^2 \int_{-\infty}^{\infty} d\omega \ e^{-i\omega
t} \Big[
D_{R}^{(1)}(\omega)\pi_{ij}(\omega)\nonumber\\&&+\frac{b^6}{2} \int_{-\infty}^{\infty}
d\omega_1 D_{R}^{(2)}(\omega, \omega_1 )
\Big[(\pi_{ik}(\omega_1)\pi_{kj}(\omega-\omega_1) +
\pi_{ik}(\omega-\omega_1)\pi_{kj}(\omega_1)
\nonumber\\&&\qquad\qquad\qquad\qquad\qquad\ \ -\frac{2}{3}\delta_{ij}\pi_{rs}(\omega_1)\pi_{rs}(\omega-\omega_1)
)\Big] + O(\delta^3)\Big] = 0,
\end{eqnarray}
where all the integrations are done with our contour
prescription which picks up contributions only from poles in the
lower half plane. Further, $D_{R}^{(2)}(\omega, \omega_1)$ is dimensionless like
$D_R^{(1)}(\omega)$ and can
be readily identified as the coefficient of
the $\log(br-1)$ term in the series expansion of
$F^{(2)}_{1}(r,\omega,\omega_1)$ at the late-time horizon
$r=1/b$.

We can also series expand $D_{R}^{(2)}(\omega,\omega_1)$
near $\omega = \omega_1 = 0$ as below,
\begin{eqnarray}\label{regseries}
D_{R}^{(2)}(\omega,\omega_1) &=&
\displaystyle\sum\limits_{n=0}^{\infty}\displaystyle\sum\limits_{\substack{m=0 \\ n+m \ is \ even}}^{n}(-ib)^n
(\omega_1^{\frac{n+m}{2}}(\omega-\omega_1)^{\frac{n-m}{2}}\nonumber\\&&\qquad\qquad\qquad\qquad\qquad+\omega^{\frac{n-m}{2}}_1(\omega-\omega_1)^{\frac{n+m}{2}})D_{R}^{(2,n,m)},
\end{eqnarray}
where $D_{R}^{(2,n,m)}$ are dimensionless numbers and the sum over $m$ runs
over even integers only if $n$ is even, and odd integers only if $n$
is odd, so that both $(n+m)/2$ and $(n-m)/2$ are integers. Comparing with \eqref{seriesef2} and recalling that $D_R^{(2)}(\omega,\omega_1)$ is the coefficient of the $\log(br-1)$ term of the series expansion of $F^{(2)}_1(r,\omega,\omega_1)$ at $r=1/b$,  we obtain that $D_{R}^{(2, n, m)}$ is the coefficient of the $\log(br-1)$ term of the series expansion of $F^{(2,n,m)}_1(r)$ at $r=1/b$. For instance, 
\begin{equation}
D_R^{(2,0,0)} = 1/2, \quad etc.
\end{equation}

Given the explicit recursion relations for $F^{(2,n,m)}_1(r)$s given by \eqref{recef} and \eqref{sol21}, we indeed obtain recursion relations for $D_R^{(2)}(\omega,\omega_1)$
implicitly. A more explicit answer, which is essentially a combinatorial question, will tell us
how we can analytically continue the series \eqref{regseries} in the lower half complex $\omega$ and $\omega_1$ planes, to obtain the regularity condition explicitly through the Fourier transform \eqref{ft2}. Unfortunately, we have not been able to achieve this here.

However, the series expansion \eqref{regseries} allows us to write the regularity
condition \eqref{ft2} up to second order in the amplitude expansion in the following form,
\begin{eqnarray}\label{rc}
C_{R \ ij}(t) &=& b^2 \displaystyle\sum\limits_{n=0}^{\infty}b^n D_R^{(1,n)}
\left(\frac{dt}{dt}\right)^n \pi_{ij} \nonumber\\
&&+b^6\displaystyle\sum\limits_{n=0}^{\infty}\displaystyle\sum\limits_{\substack{m=0 \\ n+m \ is \ even}}^{n}D_R^{(2,n,m)}b^n\displaystyle\sum\limits_{\substack{a,b=0 \\ a+b=m \\ |a-b|=m}}^{n}\Bigg[\left(\frac{d}{dt}\right)^{a}
\pi_{ik}\left(\frac{d}{dt}\right)^{b}
\pi_{kj}\nonumber\\&&\qquad\qquad\qquad\qquad  -\frac{1}{3}\delta_{ij}\left(\frac{d}{dt}\right)^{a}
\pi_{pq}\left(\frac{d}{dt}\right)^{b} \pi_{pq}\Bigg] + O(\delta^3)= 0,
\end{eqnarray} 
where again the sum over $m$ is over
even integers only when $n$ is even, and over odd integers only
if $n$ is odd, so that both $a$ and $b$ are integers. \textit{Further, the reader may recall that the above equation has been derived in the global comoving frame where the velocity of the flow vanishes. However, this equation can be also written in a Lorentz-covariant form, with $t$ replaced by $-u.x$ and $d/dt$ replaced by $u\cdot\partial$. Also the covariant form of $\pi_{ij}$ is $\pi_{\mu\nu}$, which by definition is such that its non-vanishing part is the projection to the spatial plane orthogonal to $u^\mu$, i.e. $P_\mu^{\phantom{\mu}\rho}P_\nu^{\phantom{\nu}\sigma}\pi_{\rho\sigma}$, with $P_\mu^{\phantom{\mu}\rho}$ being the spatial hyperplane projection tensor $u_\mu u^\rho + \delta_\mu^{\phantom{\mu}\rho}$. The regularity condition in gravity stated here, which is the vanishing of the $\log (br-1)$ term in the series expansion of traceless and transverse part of $G_{\mu\nu}$, is also Lorentz-covariant and therefore automatically yields this form.}

The non-linear terms in this equation of motion for
$\pi_{ij}(t)$ can have drastic effects. A specific question can be elegantly framed in the Fourier space. We know from the discussion in the previous subsection that the regularity condition at the linear order implies that $\pi_{\mu\nu}(\omega)$, when analytically continued as a function of $\omega$ in the LHP, should have an infinite series of discrete poles implying exponential decay in real time. The question is how the behavior is modified at the non-linear level, do the residues of the poles get related and the subleading $O(1/\omega^2)$ asymptotic behavior constrained by the non-linear corrections, or is the modification more radical, like appearance of branch cuts, etc.? 

We conclude our regularity analysis here by emphasizing again that the eq \eqref{rc} is valid only when the amplitude of the non-hydrodynamic $\pi_{ij}$ is small compared to the pressure, though its time derivatives can be large. Further, such solutions should be smoothly connected to the equilibrium solution, where $\pi_{ij}=0$, because the dual metric in gravity from which this eq \eqref{rc} has been obtained, is connected to the equilibrium solution by smooth variation of the amplitude perturbation parameter. This also implies that the meaningful solutions equilibrate at late-time. It thus makes no sense, for instance, to look for time-independent solutions to \eqref{rc} truncated up to second order in the amplitude expansion, though such non-trivial solutions discrete up to constant spatial rotations exist. Besides, one can also check that, for this instance, the amplitude is of O(1) and hence the perturbation expansion can not be trusted anyway.
 
\subsection{Extracting new phenomenological parameters}

In the case of the Boltzmann equation, as discussed in the Introduction, we can construct \textit{conservative solutions} \cite{myself2} for homogeneous relaxation, where we
can obtain an equation of motion for the shear-stress tensor $\pi_{ij}(t)$, such that any solution of this equation lifts to a unique solution of the full Boltzmann equation.

The shear-stress tensor $\pi_{ij}$ is actually the local symmetric and traceless rank two velocity moment tensor of $f(x,v)$, the quasiparticle distribution function. In the 
homogeneous solutions of the Boltzmann equation, the
hydrodynamic variables are constant in space and time. Further, in the conservative homogeneous solutions $\pi_{ij}(t)$ obeys an equation of motion which can be
obtained order by order in the amplitude expansion and takes the form \footnote{Here we assume that the Boltzmann equation is that of an underlying continuum system, it is not for instance a lattice Boltzmann equation. Further, the continuum system is not subject to external forces and its equilibrium state is isotropic. All these assumptions apply to gauge-theory plasmas.},
\begin{equation}\label{be}
\frac{d \pi_{ij}}{dt} = B^{(2)}(\rho, T)\pi_{ij} +
B^{(4, 2, 2)}(\rho,T)\left(\pi_{ik}\pi_{kj} + \pi_{jk}\pi_{ki} -\frac{2}{3}\delta_{ij}\pi_{lm}\pi_{lm}\right) + O(\delta^3),
\end{equation}
where $B^{(2)}$ and $
B^{(4, 2, 2)}$ can be calculated from the collision kernel of the Boltzmann equation as functions of the thermodynamic variables,
the density $\rho$ and the temperature $T$, and $O(\delta^3)$
denote terms third order in the amplitude expansion and beyond \cite{myself2}. We recall that the velocity is constant, and in a Lorentz covariant notation, $t = -u\cdot x$ and $d/dt = u\cdot\partial$. Also the covariant form of $\pi_{ij}$ is the non-vanishing part of $\pi_{\mu\nu}$, which is the projection to the spatial plane orthogonal to $u^\mu$, i.e. $P_\mu^{\phantom{\mu}\rho}P_\nu^{\phantom{\nu}\sigma}\pi_{\rho\sigma}$, with $P_\mu^{\phantom{\mu}\rho}$ being the projection tensor $u_\mu u^\rho + \delta_\mu^{\phantom{\mu}\rho}$.
In the conservative solutions, all
velocity moments of $f(x,v)$ other than the constant
hydrodynamic variables are algebraic functions of $\pi_{ij}$ and
their time derivatives, so they do not have independent dynamical parts.

The important difference of (\ref{be}) derived via the
Boltzmann equation which holds at small 't Hooft coupling, from
\eqref{rc} which holds at large 't Hooft coupling $\lambda$, is this equation 
is only first-order in time.  In fact, we know that such a Boltzmann
equation is equivalent to perturbative non-Abelian gauge
theory \cite{Arnold1, Arnold2} at sufficiently high temperature, so this is indeed a feature of  perturbative non-equilibrium non-Abelian gauge theories. The higher order time derivatives in (\ref{rc})
therefore have their origins in the intrinsic energy-time
uncertainty of quantum dynamics going beyond the semiclassical
approximation in the Boltzmann equation, according to which we
cannot define the energy of the excitation at short time scales.
As of now, we do not know how to capture these effects in non-equilibrium non-Abelian
gauge theories systematically.

In case of a general inhomogeneous conservative solution, $\pi_{ij}(x,t)$ can be
split as $\pi_{ij}^{(nh)}(x,t)+\pi_{ij}^{(h)}(x,t)$. Here,
$\pi_{ij}^{(h)}$ is purely hydrodynamic and is given as an algebraic
function of the hyrodynamic variables and their spatial
derivatives in the derivative expansion (in relativistic case, this is true only in a local
inertial frame where the local mean velocity of the quasiparticles vanishes). On the other hand $\pi_{ij}^{(nh)}$ satisfies a equation of
motion which is first order in time derivative (in the
relativistic case only in the mentioned local inertial frame), and can be expanded in both the amplitude and derivative expansions. Both can be obtained systematically from the Boltzmann equation. In both cases, the derivative expansion refers to the spatial derivatives also in a local inertial frame where the local mean velocity vanishes.

Because the Boltzmann equation itself is first-order in
time, all higher time
derivatives can be eliminated in favour of spatial derivatives
in the conservative solutions, in an appropriate local inertial frame just mentioned, so that the equation of motion of the hydrodynamic variables and the shear-stress tensor
are first order in time. All other velocity moments of $f(x,v)$ are algebraic
functions of the hydrodynamic variables, the shear-stress tensor and their spatial derivatives,
in the same local inertial frame, so do not have independent dynamical parts.

The amplitude and derivative expansion become complicated if we
go beyond the Boltzmann approximation. The 
conservative solutions should also exist beyond this approximation, 
as at large $N$ and 't
Hooft coupling $\lambda$ this would naturally explain why in
the supergravity approximation the universal sector of
non-equilibrium states can be determined from the dynamics of
energy-momentum tensor alone. This has already been discussed in
the Introduction. However, there is no reason why the equations
of motion for hydrodynamic variables and $\pi_{ij}$ should be only
first-order in time in a local inertial frame where the
mean velocity vanishes.

In presence of higher order time derivatives in the kinetic equation, we cannot treat
the time derivatives of the non-hydrodynamic $\pi_{ij}^{(nh)}$ perturbatively. Though $\pi_{ij}^{(nh)}$ becomes arbitrarily
small compared to the pressure close to equilibrium, its
time derivatives do not become small with respect to average
time between quasiparticle collisions even close to equilibrium, as
it decays exponentially at linear order. This means that we
should treat the derivatives of hydrodynamic variables
perturbatively (including time derivatives), but we can treat
the amplitude of $\pi_{ij}^{(nh)}$ perturbatively only if we sum
over all its time derivatives. However, close to equilibrium, we can consider a regime where the spatial derivatives are small.

At large 't Hooft coupling $\lambda$ and large $N$, we can expect that
the conservative solutions will
form the universal sector of non-equilibrium states dual to the
solutions of pure gravity with regular future horizons. In that
case, we get the desired equation of motion of the
energy-momentum tensor from the regularity condition on the late-time
horizon in the bulk metric. In the purely hydrodynamic case before, and here in the
case of homogeneous relaxation, such equations of motion have
been thus obtained from gravity. However, we can also write down a phenomenological
equation for the energy-momentum tensor which will hold for an abitrary configuration which equilibrates in the future.
In absence of a known Boltzmann-like equation or any
well-defined reliable formalism in the strongly coupled regime,
the only guide to such a phenomenological equation is conformal
covariance.

In \cite{myself2}, we did propose such a phenomenological equation for the
energy-momentum tensor, such that for the right values of the
dimensionless phenomenological coefficients, it should reproduce all
solutions of gravity with regular future horizons which
equilibrate at late-time. However, we did not realize that we
need to sum over all time derivatives of the non-hydrodynamic
shear-stress tensor at each order in the
amplitude expansion. This did not allow us to
compare with quasinormal modes correctly. We rectify this here
and see how the analysis of homogeneous relaxation
fixes some of the general phenomenological coefficients.

To begin with, we will split the relativistic shear-stress tensor
$\pi_{\mu\nu}$ into a purely hydrodynamic part $\pi^{(h)}_{\mu\nu}$ algebraically determined by hydrodynamic variables and $\pi^{(nh)}_{\mu\nu}$ whose evolution has dependence on both hydrodynamic and non-hydrodynamic parameters, and is therefore independent of the hydrodynamic variables dynamically. The amplitude
expansion is concerned with only the non-hydrodynamic part, the
dimensionless expansion parameter being its ratio with the
pressure at late-time equilibrium. The hydrodynamic
$\pi^{(h)}_{\mu\nu}$ is by definition an algebraic functional of
the hydrodynamic variables and their derivatives (even
time derivatives). The derivative expansion is concerned with
spatial and temporal rate of variation of the hydrodynamic
variables and only spatial rate of variations of
$\pi^{(nh)}_{\mu\nu}$, compared to the temperature at late-time
equilibrium. Note the spatial variation of even $\pi^{(nh)}_{\mu\nu}$ is small close to equilibrium, but its time derivative is not.

Time-derivatives involve $(u\cdot\partial)$,
however if this acts on a Weyl-covariant quantity like
$\pi_{\mu\nu}^{(nh)}$, the resulting quantity will not be
Weyl-covariant. So, we need to define a Weyl-covariant form of
the time derivative $\mathcal{D}$, following \cite{Loganayagam}, whose action on a
Weyl-covariant tensor $\mathcal{A}_{\mu\nu}$ turns out to be
\begin{equation}
\mathcal{D}\mathcal{A}_{\mu\nu} =
(u\cdot\partial)\mathcal{A}_{\mu\nu} +
\frac{4}{3}\mathcal{A}_{\mu\nu}(\partial\cdot u) -
\left(\mathcal{A}_{\mu} ^{\phantom{\mu}\beta} u_{\nu} +
\mathcal{A}_{\nu} ^{\phantom{\nu}\beta} u_{\mu}\right) (u \cdot
\partial)u_{\beta}.
\end{equation}
Therefore, more precisely by time derivatives of $\pi_{\mu\nu}^{(nh)}$ we will denote successive actions of
the Lorentz invariant and Weyl-covariant derivative $\mathcal{D}$. The spatial
derivatives will denote actions of
$P_{\mu}^{\phantom{\mu}\nu}\partial_{\nu}$, where
$P_{\mu}^{\phantom{\mu}\nu}$ as before denotes $u_{\mu}u^{\nu}+
\delta_{\mu}^{\phantom{\mu}\nu}$, the projection on the spatial
plane orthogonal to $u_{\mu}$.

The conservation of energy-momentum tensor implies
\begin{equation}\label{ce}
\partial^{\nu}\left(\frac{3u_\mu u_\nu +P_{\mu\nu}}{4b^4} +
\pi_{\mu\nu}^{(h)} + \pi_{\mu\nu}^{(nh)}\right) = 0,
\end{equation}
where $b=1/(\pi T)$ as before, and all quantities are dependent
on both space and time coordinates. We also use the earlier
field definitions of $T$ and $u^{\mu}$, such that $(3/4)(\pi
T)^4$ is the local energy density and $u^{\mu}$ is the local
four-velocity of energy-transport so that
$u^{\mu}\pi_{\mu\nu}^{(h)} = u^{\mu}\pi_{\mu\nu}^{(nh)} = 0$.
The traceless-ness of the energy-momentum tensor also implies
that $Tr(\pi^{(h)})= Tr(\pi^{(nh)})= 0$. With these conditions,
$\pi_{\mu\nu}^{(nh)}$ has five independent components only. As
$\pi_{\mu\nu}^{(h)}$ is an algebraic function of $u^{\mu}$ and
$T$, and their derivatives, the nine independent variables
parametrising the energy-momentum tensor are $T$, the three
independent variables of the four velocity $u^{\mu}$ and the
five independent components of $\pi_{\mu\nu}^{(nh)}$.

The form of the hydrodynamic part of the shear-stress tensor can be
determined by Weyl covariance and the transport coefficients can be fixed by the regularity condition at the future horizon,
order by order in the derivative expansion. Up to second order
in the derivative expansion, we obtain
\begin{eqnarray}\label{soh}
\pi_{\mu \nu}^{(h)} &=& - \frac{2}{b^3} \sigma_{\mu \nu} +
\frac{2 - \log 2}{b^2} \mathcal{D}\sigma_{\mu \nu} \nonumber\\
&&+ \frac{2}{b^2}
\left(\sigma_{\mu}^{\phantom{\mu}\alpha}\sigma_{\alpha\nu} -
\frac{1}{3}P_{\mu \nu} \sigma_{\alpha \beta}\sigma^{\alpha
\beta}\right) \nonumber\\
&&+\frac{\log 2}{b^2}
(\sigma_{\mu}^{\phantom{\mu}\alpha}\omega_{\alpha\nu}
+\sigma_{\nu}^{\phantom{\nu}\alpha}\omega_{\alpha\mu} ) +
O(\epsilon^3),
\end{eqnarray}
where $O(\epsilon^3)$ denotes terms higher order in the
derivative expansion (where the expansion parameter $\epsilon$ is the typical length scale of variation with respect to the temperature), and
\begin{eqnarray} 
\sigma_{\mu\nu} &=&
\frac{1}{2}P_{\mu}^{\phantom{\mu}\alpha}P_{\nu}^{\phantom{\nu}\beta}\left(\partial_{\alpha}u_{\beta}+\partial_{\beta}u_{\alpha}\right)
-\frac{1}{3} P_{\mu\nu}(\partial \cdot u), \nonumber\\
\omega_{\mu\nu} &=&
\frac{1}{2}P_{\mu}^{\phantom{\mu}\alpha}P_{\nu}^{\phantom{\nu}\beta}(\partial_{\alpha}u_{\beta}-\partial_{\beta}u_{\alpha}).
\end{eqnarray}

The equation of motion of $\pi_{\mu\nu}^{(nh)}$ should be such
that it is (a) Weyl-covariant and
(b) we can consistently put $\pi_{\mu\nu}^{(nh)}=0$, so that we
can go to the pure hydrodynamic limit. In the case of the
conservative solutions of the Boltzmann equation, this is always
possible. Even in the case of gravity, we can construct
solutions which are regular order by order in the
derivative expansion when $\pi_{\mu\nu}^{(nh)}= 0$, hence requirement (b) is 
essential.

With these two requirements we can write the most general
equation of motion for $\pi_{\mu\nu}^{(nh)}$ as
\begin{eqnarray}\label{rcg}
\left(\displaystyle\sum\limits_{n=0}^{\infty}D_{R}^{(1,n)}\frac{1}{b^n}
\mathcal{D}^n\right)\pi_{\mu \nu}^{(nh)} &=&
\frac{\lambda_1 }{2b}
\left(\pi_{\mu}^{(nh)\alpha}\sigma_{\alpha\nu}
+\pi_{\nu}^{(nh)\alpha}\sigma_{\alpha\mu}- \frac{2}{3}P_{\mu
\nu} \pi_{\alpha \beta}^{(nh)}\sigma^{\alpha \beta}\right)
\\\nonumber
&&+\frac{\lambda_2 }{2b}
\left(\pi_{\mu}^{(nh)\alpha}\omega_{\alpha\nu}
+\pi_{\nu}^{(nh)\alpha}\omega_{\alpha\mu}\right) \\\nonumber
&&- \frac{1}{b^4}\displaystyle\sum\limits_{n=0}^{\infty}\displaystyle\sum\limits_{\substack{m=0 \\ n+m \ is \ even}}^{n}D_R^{(2,n,m)}\frac{1}{b^n}\displaystyle\sum\limits_{\substack{a,b=0 \\ a+b=n \\ |a-b|=m}}^{n}\Bigg[\mathcal{D}^a
\pi_{\mu}^{(nh)\alpha}\mathcal{D}^b
\pi_{\alpha\nu}^{(nh)}\\\nonumber &&\qquad\qquad\qquad\qquad\qquad\qquad\qquad  -\frac{1}{3}P_{\mu\nu} \ \mathcal{D}^{a}
\pi_{\alpha\beta}^{(nh)}\mathcal{D}^b \pi^{(nh)\alpha\beta}\Bigg]\\\nonumber
&&+ O(\epsilon^2\delta, \epsilon\delta^2 ,\delta^3).
\end{eqnarray}
Here, we note that the RHS is at least linear in the amplitude
expansion (where the expansion parameter $\delta$ is the ratio of the typical non-hydrodynamic shear-stress with respect to the equilibrium pressure), so that $\pi^{(nh)}$ can be put to zero consistently.
Further we have included terms only up to first order in
derivative expansion on the RHS. The terms explicitly shown in
RHS are thus $O(\epsilon\delta)$ and $O(\delta^2)$. We cannot
include any term like $\partial^{\mu}\pi_{\mu\nu}^{(nh)}$ here
because by (\ref{ce}) it gets related to purely hydrodynamic
quantities and their derivatives. By definition, all the
phenomenological parameters $\lambda_1$, $\lambda_2$,
$D_R^{(1,n)}$ and $D_R^{(2,n,m)}$ are dimensionless. Further
the condition $n+m$ is even in the summation, denotes that sum over 
$m$ is over odd integers only when $n$ is odd
and is over even integers only when $n$ is even, so that both
$(n+m)/2$ and $(n-m)/2$, i.e. $a$ and $b$, are integers. This phenomenological
equation is the right equation to consider so that we have
indeed summed over all time derivatives of the non-hydrodynamic
shear-stress tensor correctly.

This equation (\ref{rcg}), together with (\ref{ce}) and
(\ref{soh}) give the complete equations of motion for the
energy-momentum tensor in the general case. Further for the
right choices of the phenomenological coefficients in
(\ref{rcg}), the dual solution in gravity should have regular
future horizons.

We note that when $u^{\mu}$ and $T$ are constants in space and
time, and the fluid is at rest, the equation for conservation of energy and momentum given by
(\ref{ce}) implies that
$\pi_{00}^{(nh)}=\pi_{0i}^{(nh)}=\pi_{i0}^{(nh)} =0$, and
$\pi_{ij}^{(nh)}$ is a function of time only. Also, in such a
case, the Weyl-covariant derivative $\mathcal{D}$ reduces to an
ordinary time derivative. We can readily see that the equation
(\ref{rcg}) rduces to (\ref{rc}), so that we recover the
regularity condition at the late-time horizon for metrics dual
to homogeneous relaxation.

We cannot determine the phenomenological coefficients $\lambda_1$ and $\lambda_2$ appearing in \eqref{rcg} by considering homogeneous configurations as we have done here, because the corresponding terms involving derivatives of the velocity like $\sigma_{\mu\nu}$ and $\omega_{\mu\nu}$ vanish, as the velocity remains constant. We can determine these coefficients by considering inhomogeneous configurations and adapting our method, provided we can argue that the metric should be regular in the ingoing Eddington-Finkelstein coordinates.

\section{Outlook for RHIC and ALICE } 

We briefly mention here the relevance of this work for the great challenge of modelling the space-time evolution of the matter formed by ultra-relativistic heavy ion collisions at RHIC and ALICE.  Experiments at RHIC suggest the validity of the following picture \cite{Florkowski} : (i) a large fraction of the initial kinetic energy of the colliding ions is thermalized astonishingly fast (in time $\leq$ 1 fm) forming a locally equilibrated hot and dense fireball parametrized by a profile of the hydrodynamic variables - namely the temperature, four-velocity and chemical potential fields, (ii) the strongly interacting fireball undergoes hydrodynamic expansion which can be described by the hydrodynamic equations like the Navier-Stokes' equation, and (iii) the initial transverse hydrodynamic flow at the time of local thermalization in most cases vanishes. Most of the data at RHIC is in good agreement with this simplistic picture especially in the mid-rapidity region, i.e. for the most central collisions at the highest beam energy of $\sqrt{s_{NN}} = 200$ GeV. However, despite the success in explaining the transverse momentum spectra of hadrons, the elliptic flow coefficient, etc., this does not reproduce pion interferometric data like HBT radii leading to the well-known RHIC HBT puzzle.

It is necessary to have a better phenomenological model for the space-time evolution of the fireball to explain the data completely, and also to reduce theoretical uncertainties. The best theoretical tool at hand for studying evolution of strongly coupled matter of gauge theories in real time is the AdS/CFT correspondence. It is well-known that the AdS/CFT correspondence gives $\eta/s = 1/4\pi$ \cite{Policastro1}, while the current analysis of experimental data suggests $1 < 4\pi\Big(\eta/s\Big) < 2.5$ for temperatures probed at RHIC \cite{Heinz}. 

Here we will propose that the AdS/CFT correspondence can be used to develop a complete phenomenology for the evolution of the strongly coupled matter, describing both the late stages of local thermalization and the subsequent hydrodynamic expansion in an unified framework. These phenomenological equations involve a closed set of equations for evolution of the energy-momentum tensor and the baryon number  charge current alone. For simplicity, if we assume  that the baryon number chemical potential is zero throughout the evolution, then the full set of phenomenological equations are \eqref{ce}, \eqref{soh}, \eqref{rcg}. It should be straightforward to include the dynamics of the charge current but we will not attempt this here.

The advantage of our proposal is that there is a very natural way to connect the expansion of the fireball with any model which describes the early stages of the collision process, as for instance the parton cascade model \cite{Geiger}. All that is needed is to match the evolution of the energy-momentum tensor and conserved charge currents before and after the matter enters in the strongly coupled phase of evolution. Importantly, the matching with the initial regime does not require the energy-momentum tensor to be hydrodynamic.

To state more concretely, the space-time evolution of the matter till chemical and kinetic freezeout with subsequent hadronization can be divided into two phases :

\begin{itemize}
\item{\textbf{The entry into strongly coupled phase :}} This phase describes the very initial stage of particle production at high energies and can probably be described in the perturbative framework as in the parton cascade model \cite{Geiger}. The energy-momentum tensor can be computed by such models and the local temperature fields can be obtained from the Landau-Lifshitz decomposition of the energy-momentum tensor. As mentioned before, any arbitrary energy-momentum tensor can be written in the Landau-Lifshitz form and thus we can always obtain the local effective temperature. However, this requires using an appropriate equation of state for the initial phase. The effective temperature will automatically give us when the matter locally enters the strongly coupled phase. A more detailed analysis of this initial phase is beyond the scope of this paper.

\item{\textbf{The fireball formation and its expansion :}} Once the matter enters into the strongly coupled phase, we can use the phenomenological equations from AdS/CFT to describe both the local thermalization and late hydrodynamic expansion in an unified framework. The initial conditions can be obtained by matching the energy-momentum tensor with the model describing the initial stage \footnote{This  matching is however complicated as the equations of evolution involve time derivatives of arbitrary orders.}. As the matter is expanding and boost invariance is a good approximation for central collisions, we may use the boost invariant versions of eqs. \eqref{ce}, \eqref{soh}, \eqref{rcg}.

Homogeneous relaxation in the boost-invariant regime which estimates the time for local thermalization or fireball formation has been studied in the linearized approximation in \cite{Janikhom}.  Furthermore, as also shown in \cite{Janikhom}, the homogeneous quasinormal modes in the static case can be mapped readily to boost invariant geometries. The boost-invariant version of our equations, as discussed here, must reproduce these linearized fluctuations of the boost-invariant geometry which are homogeneous in transverse coordinates. The lowest static homogeneous quasinormal mode gives an estimate of the time for local equilibration, i.e. fireball formation. In the boost invariant case, the time-dependence in the linearized plane wave approximation is not exponential, but given by a Bessel function as in $\sqrt{\tau} J_{\pm \frac{3}{4}}\Big(\frac{3}{2} \omega \tau^{\frac{2}{3}} (\pi T_f)^{-\frac{1}{3}}\Big)$, where $\omega \approx 2.74667 (\pi T_f)$ and $T_f$ is the fireball temperature which is about 175 MeV at RHIC. At late time, this reduces to a proper-time damping of the form : $exp \Big(-\frac{3}{2}\cdot \ 2.74667\cdot \ (\pi T_f \tau)^{\frac{2}{3}}\cdot \Big)$.

In future work, we would like to study the boost-invariant version of the non-linear homogeneous equations given by \eqref{rcg} and see numerically how this behavior both qualitatively and quantitatively changes due to non-linearities. Further, it should be also possible to include inhomogeneities in transverse directions and also study the transition of the fireball to the hydrodynamic regime using the full set of equations - \eqref{ce}, \eqref{soh}, \eqref{rcg}. In practice, one needs to use these equations in a coordinate system (as in \cite{Janik1}) better adapted for the late equilibrium state of the fireball which is an ideal fluid undergoing boost-invariant expansion. This coordinate system comprises of the proper time coordinate $\tau$ of the late time expansion, the coordinate $y$ parameterizing rapidity, and the two transverse coordinates $x^1$ and $x^2$. 

\end{itemize}

\section{Discussion}

 From the point of view of gauge/gravity duality, our work leads a few open questions. To conclude, we will discuss two of the most immediate and important of them here.
\begin{itemize}
\item{\textbf{Is there an entropy current for homogeneous relaxation?}} In the case of pure hydrodynamic behavior, it is known that the horizon in the bulk metric also gets deformed mildly, and the deformation can be calculated order by order in the derivative expansion. This has been shown in \cite{Bhattacharyya2}, where it has been also demonstrated that the pull-back of the Hodge-dual of the area form on the horizon world-volume to the boundary gives the construction of a family of entropy currents whose divergences are positive-definite, order by order in the derivative expansion. 

This is an interesting finding, because to our knowledge, this gives the first construction of an 
exact entropy current in the hydrodynamic regime, beyond any approximation like the limit of validity of the Boltzmann equation. An interesting question is whether gauge/gravity duality implies the existence of such entropy currents at large $N$, even beyond the hydrodynamic regime.

In our case of spatially homogeneous relaxation, even with the explicit metrics obtained here, this is a question which is not easy to answer. This is because the location of the horizon $r(t)$ as a function of time will fluctuate only perturbatively in the amplitude expansion, but its time derivatives will not be under control. Since we have summed the time derivatives only recursively here, this does not reveal the global nature of the horizon even perturbatively in the amplitude expansion.  We would have to tackle this question possibly numerically later.

Alternatively, we can also analyze the general phenomenological eqs. \eqref{ce}, \eqref{soh} and \eqref{rcg} and find conditions under which these imply existence of entropy current(s). Then, one can verify if these conditions are satisfied by the phenomenological coefficients obtained by gauge/gravity duality.

\item{\textbf{What are the field-theoretic definitions of the general phenomenological coefficients?}} Recently, it has been possible to obtain a systematic procedure for constructing Kubo-like formulae for all hydrodynamic transport coefficients (including those which appear only non-linearly) in terms of the low frequency and large wavelength expansion of $n$-point correlation functions of the energy-momentum tensor \cite{Moore}. It has so far not been possible to define such general formulae for non-hydrodynamic parameters beyond any approximation scheme.

Our phenomenological coefficients (including those involved in homogeneous relaxation), can be in principle obtained by constructing \textit{conservative solutions} of non-equilibrium field theory. However, we have been able to construct these solutions in \cite{myself2} in the perturbative regime in non-Abelian gauge theories only. It would be interesting to see if we can also find a non-perturbative method of constructing these conservative solutions and relate the phenomenological coefficients to $n$-point correlation functions of the energy-momentum tensor.

If the above is possible, it would also provide a non-trivial consistency check of gauge/gravity
duality. The $n$-point correlation functions of the energy-momentum tensor can be independently calculated by gauge/gravity duality, so we can obtain the phenomenological coefficients through them and see if they match with the values required by the regularity condition on the future horizon. 

\item{\textbf{Is the ingoing Eddington-Finkelstein coordinate system the right choice for all configurations which equilibrate?}} It would be certainly of interest to see if the bulk metric is manifestly regular at the future horizon in the ingoing Eddington-Finkelstein coordinate system, for general configurations which equilibrate at late time. If this is so, we can readily extend our proof to show that the general regularity condition for the future horizon in the bulk, is indeed given by \eqref{ce}, \eqref{soh} and \eqref{rcg}, for the right values of the
phenomenological coefficients.
 
\end{itemize}
\textbf{Acknowledgments :} AM thanks G. Policastro, Ashoke Sen, R. Gopakumar, M. Paulos and Iosif Bena for motivating discussions over the course of this work. AM acknowledges the organizers of the Indian String Meeting, Puri, January 2011, for the opportunity to discuss aspects of this work with some of the participants. AM also thanks HRI, Allahabad, India for the opportunity to present this work while it was being completed, and IPHT, CEA-Saclay and Institute Henri Poincar\'{e}, Paris for inviting to present this work before it has appeared in the present version. The research of AM is supported by the grant number ANR-07-CEXC-006 of the Agence Nationale de La Recherche.

\appendix
\section{The Fefferman-Graham vector and scalar equations at the second order in the amplitude expansion}
The vector eq. \eqref{ecv} at the second order in the amplitude expansion reduces to
\begin{eqnarray}\label{eom2v}
\omega \left(V_{l}(\rho)  + 3 \frac{\partial}{\partial \rho}\right) f^{(2)}_2 (\rho,\omega,\omega_1) &=& 
\omega V_{1r} (\rho) f^{(1)}(\rho,\omega_1)f^{(1)}(\rho,\omega - \omega_1) \nonumber\\
&& + V_{2r}(\rho) (\omega + \omega_1) f^{(1)}(\rho, \omega - \omega_1) \frac{\partial}{\partial \rho}f^{(1)}(\rho,\omega_1)\nonumber\\
&&+ V_{2r}(\rho) (2\omega - \omega_1) f^{(1)}(\rho, \omega_1) \frac{\partial}{\partial \rho}f^{(1)}(\rho,\omega - \omega_1) ,
\end{eqnarray}
where
\begin{eqnarray}
V_l(\rho) = \frac{12 \rho^3}{(4b^4 - \rho^4)}, \quad V_{1r}(\rho) = \frac{4 \rho^3 b^4 (4b^4 + 3 \rho^4)}{(4b^4 - \rho^4)(4b^4 + \rho^4)^2}, \quad V_{2r}(\rho) = \frac{b^4}{(4b^4 + \rho^4)}.
\end{eqnarray}
A close inspection shows that the right hand side of \eqref{eom2v} is symmetric under the exchange of $\omega_1$ with $\omega-\omega_1$ as it should be.

The scalar eq. \eqref{ecs} at the second order in the amplitude expansion reduces to
\begin{eqnarray}\label{eom2s}
&&\left(S_{1l}(\rho) + S_{2l}(\rho)\frac{\partial}{\partial \rho} + S_{3l}(\rho)\frac{\partial^2}{\partial \rho^2}\right)f_2^{(2)}(\rho, \omega,\omega_1) 
\nonumber\\&&+\left(S_{4l}(\rho)+ S_{5l}(\rho)\frac{\partial}{\partial \rho} + S_{6l}(\rho)\frac{\partial^2}{\partial \rho^2}\right)f_3^{(2)}(\rho, \omega,\omega_1) \nonumber\\&&= S_{1r}(\rho) f^{(1)}(\rho, \omega_1)f^{(1)}(\rho, \omega - \omega_1) + S_{2r}(\rho) \frac{\partial}{\partial \rho} f^{(1)}(\rho,\omega_1) \frac{\partial}{\partial \rho}f^{(1)}(\rho,\omega - \omega_1)\nonumber\\
&&+ S_{3r}(\rho)\left( f^{(1)}(\rho, \omega_1) \frac{\partial}{\partial \rho}f^{(1)}(\rho, \omega - \omega_1) +  f^{(1)}(\rho, \omega - \omega_1) \frac{\partial}{\partial \rho}f^{(1)}(\rho, \omega_1)\right)\nonumber\\
 &&+ S_{4r}(\rho)\left( f^{(1)}(\rho, \omega_1) \frac{\partial^2}{\partial \rho^2}f^{(1)}(\rho,\omega - \omega_1) +  f^{(1)}(\rho,\omega - \omega_1) \frac{\partial^2}{\partial \rho^2}f^{(1)}(\rho,\omega_1)\right) ,
\end{eqnarray}
where,
\begin{eqnarray}
S_{1l}(\rho) &=& \frac{24\rho^2(4b^4-\rho^4)}{(4b^4 + \rho^4)^3}, \quad S_{4l}(\rho) = \frac{8\rho^2(192b^{12} + 48\rho^4b^8 + 36 \rho^8b^4 + r^{12})}{(4b^4 - \rho^4)^4(4b^4 + \rho^4)}, \nonumber\\
S_{2l}(\rho) &=& \frac{3(4b^4 + 5\rho^4)}{\rho(4b^4 + \rho^4)^2}, \quad S_{5l}(\rho) = \frac{(-16b^8 + 48 \rho^4 b^4 + 5 \rho^8)}{\rho(4b^4 - \rho^4)^3}, \nonumber\\
S_{3l}(\rho) &=& -\frac{3}{(4b^4 + \rho^4)}, \quad S_{6l}(\rho) = \frac{(4b^4 + \rho^4)}{(4b^4 - \rho^4)^2}.
\end{eqnarray}

\end{document}